\documentclass[prb,twocolumn,reprint,graphicx,showpacs,superscriptaddress,longbibliography]{revtex4-1}
\usepackage{amsmath}
\usepackage{amssymb}
\usepackage{amsfonts}
\usepackage{mathptmx}
\usepackage{dcolumn}
\usepackage{eucal}
\usepackage{bm}
\usepackage{units}
\usepackage{color}
\usepackage[colorlinks,linkcolor=blue,citecolor=blue]{hyperref}

\setcitestyle{numbers,square}

\usepackage{graphicx}

\usepackage{natbib}

\newcommand{\PV}[1]%
   {\begingroup{\color{cyan}\it (PV: #1)}\endgroup}

\newcommand{\FG}[1]%
   {\begingroup{\color{red}\it (FG: #1)}\endgroup}

\newcommand{\AR}[1]%
   {\begingroup{\color{blue}\it (AR: #1)}\endgroup}

\DeclareMathOperator{\artanh}{artanh}
\DeclareMathOperator{\arcosh}{arcosh}
\DeclareMathOperator{\sgn}{sgn}
\renewcommand{\Re}{\mathop{\mathrm{Re}}}
\renewcommand{\Im}{\mathop{\mathrm{Im}}}

\begin{document}
\title{Josephson photodetectors via temperature-to-phase conversion}

\author{P. Virtanen}
\email{pauli.virtanen@nano.cnr.it}
\affiliation{NEST, Instituto Nanoscienze-CNR and Scuola Normale Superiore, I-56127 Pisa, Italy}

\author{A. Ronzani}
\affiliation{Low Temperature Laboratory, Department of Applied Physics, Aalto University, 00076 Aalto, Finland}

\author{F. Giazotto}
\email{francesco.giazotto@sns.it}
\affiliation{NEST, Instituto Nanoscienze-CNR and Scuola Normale Superiore, I-56127 Pisa, Italy}

%\pacs{85.80.Lp,74.50.+r,72.25.-b}
\begin{abstract}
 We theoretically investigate the \emph{temperature-to-phase}
 conversion (TPC) process occurring in dc superconducting quantum
 interferometers based on superconductor--normal metal--superconductor
 (SNS) mesoscopic Josephson junctions.  In particular, we predict the
 temperature-driven rearrangement of the phase gradients in the
 interferometer under the fixed constraints of fluxoid quantization
 and supercurrent conservation.  This allows sizeable phase variations across the
 junctions for suitable structure parameters and temperatures.  We
 show that the TPC can be a basis for sensitive single-photon sensors
 or bolometers. We propose a radiation detector realizable with
 conventional materials and state-of-the-art nanofabrication
 techniques.  Integrated with a superconducting quantum interference
 proximity transistor (SQUIPT) as a readout set-up, an aluminum
 (Al)-based TPC calorimeter can provide a large signal to noise (S/N)
 ratio $>100$ in the 10~GHz$\cdots$10~THz frequency range, and a
 resolving power larger than $10^2$ below 50~mK for THz photons.  In
 the bolometric operation, electrical NEP of
 $\sim10^{-22}$~W$/\sqrt{\text{Hz}}$ is predicted at 50
 mK.  This device can be attractive as a cryogenic single-photon sensor
 operating in the giga- and terahertz regime, with applications
 in dark matter searches.
\end{abstract}

\maketitle

\section{Introduction}

\begin{figure}
  \includegraphics[width=\columnwidth]{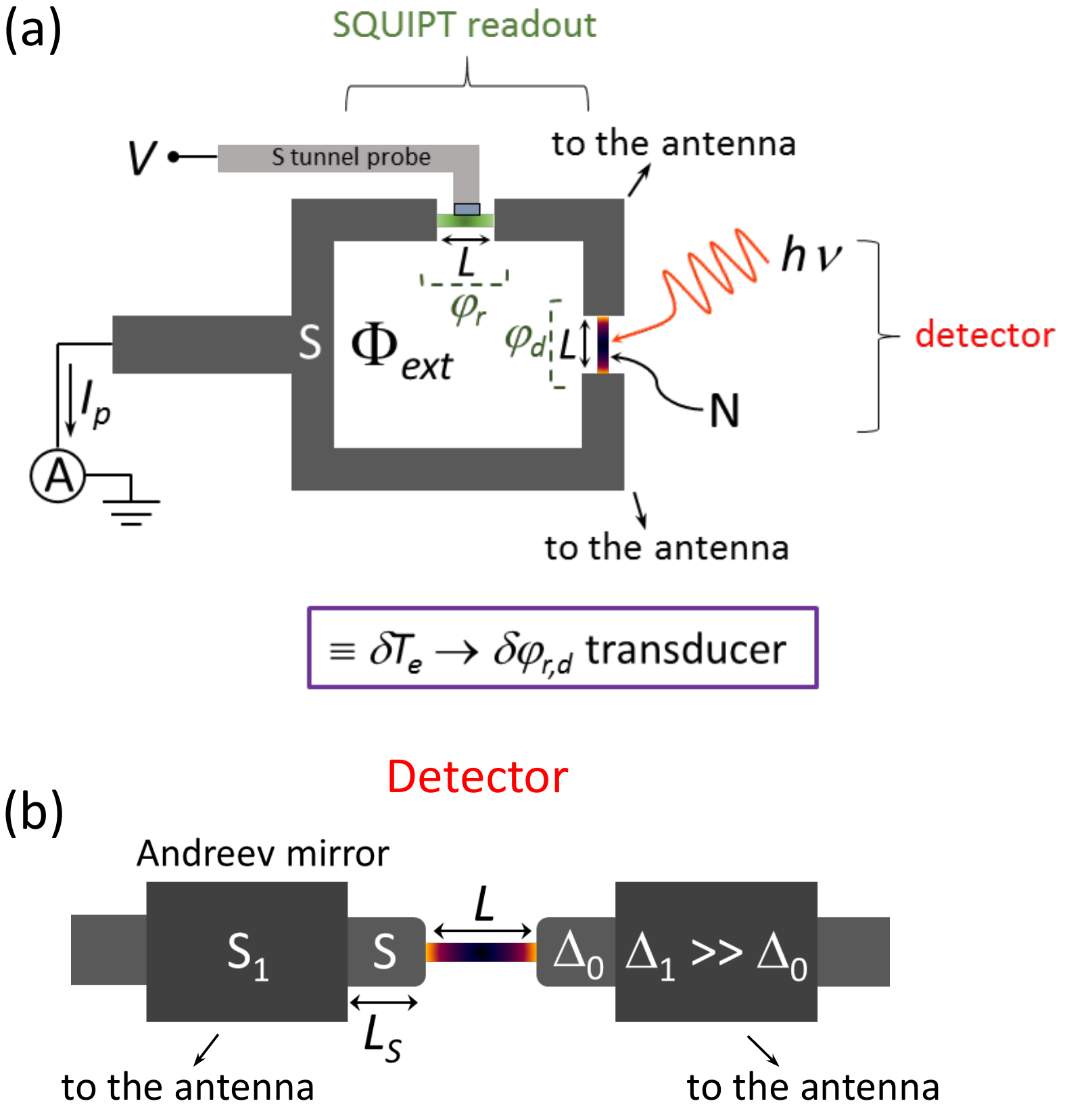}\vspace{-3mm}
  \caption{(a) Scheme of a dc superconducting quantum interferometer
    realized with two SNS Josephson junctions.  Absorption of
    electromagnetic radiation of energy $h\nu$ increases the
    electronic temperature ($T_e$) in the N region of the detector
    junction thereby suppressing the Josephson supercurrent
    circulating in the loop. This leads to a variation of phase
    difference ($\varphi_{d,r}$) across both Josephson junctions, and
    to a modification of the DOSs in their N regions. The DOSs in the
    readout junction is probed via a superconducting quantum
    interference proximity transistor (SQUIPT) used as the readout
    circuit.  The structure therefore operates as a
    temperature-to-phase converter, i.e., $\delta T_e\rightarrow
    \delta \varphi_{d,r}$. $L$ denotes the length of the junctions,
    and $\Phi_{ext}$ is the applied magnetic flux.
    (b) Blow-up of the
    detector in the vicinity of the weak-link region: Superconducting
    electrodes S$_1$ with energy gap $\Delta_1\gg\Delta_0$ ($\Delta_0$
    is the zero-temperature energy gap in S) in good electric contact
    with the loop constitute Andreev mirrors for the energy absorbed
    by quasiparticles in the N region.  This allows energy relaxation
    in the detector junction to occur predominantly via
    electron-phonon interaction with lattice phonons in the enclosed N and S
    regions (see text). $L_S$ is the length of the S portions in
    direct contact with the N nanowire.
    \label{fig:setup}}
\end{figure}

Detecting radiation in the microwave to terahertz regime has a range
of applications from astrophysics to quantum devices.  For
applications where cooling to low temperatures is available,
superconducting sensors provide a way to construct high-sensitivity
detectors, both for bolometric detection of continuous power,
\cite{semenov2002-hee,karasik2011-ntp,giazotto2006-omi,wei2008-uhe} and also towards detection of
few photons.
\cite{cabrera1998-dsi,karasik2012-erd,kerman2006-kil,govenius2016-dzm,santavicca2010-ert}

Thermal superconducting radiation detectors
\cite{bluzer1994-sqd,semenov2002-hee,giazotto2006-omi,karasik2011-ntp} generally make
use of the sensitive dependence of superconducting electrical response
on the local electronic temperature, which changes as photons are
absorbed. The specific way this is utilized varies in different
detectors. Kinetic inductance
\cite{day2003-bsd,semenov2002-hee,mazin2009-mki} and Josephson
superconductor--normal--superconductor (SNS)
\cite{giazotto2008-upj,govenius2016-dzm,govenius2014-mnb,voutilainen2010-ppj}
based detectors make use of the temperature dependence of the
supercurrent.  The changes in the current response can be detected
e.g. via resonant circuits \cite{day2003-bsd} or inductively or
directly \cite{bluzer1995-aqs,semenov2002-hee} coupled SQUID
sensors. In the direct coupling scheme, variation of the critical
current or inductance in the detector embedded as one junction of a dc
SQUID affects the distribution of phase differences across the second
junction and the current balance, which can be detected.  Similarly,
if the two junctions reside in a closed superconducting ring, the
result is \emph{temperature to phase difference conversion} (TPC),
which is the basis of the device discussed below.

The phase difference across a superconducting junction can be
generally determined via the current flowing through it. However, in
SNS and weak links, it is also possible to determine the value via tunnel junction
spectroscopy --- this forms the basis for the superconducting quantum
interference proximity transistor (SQUIPT).  \cite{giazotto2010-sqi}
In a SQUIPT, the phase difference is measured by observing the
tunneling current through a tunnel junction, \cite{giazotto2006-omi}
connected to the middle of the weak link. Such measurements
can also be made at high bandwidth,
\cite{gasparinetti2015-fet,saira2016-dtj} enabling fast measurement timescales
that are required for calorimetry.

Superconducting weak links are useful in hot-electron bolometers (HEB)
in that the superconductivity inhibits electronic heat conduction out
of the detector region,
\cite{nahum1993-uhe,semenov2002-hee,karasik2011-ntp} improving
sensitivity via reduction of the intrinsic thermal fluctuation noise.
For bolometric devices, another key quantity is the detector heat capacity
\cite{semenov2002-hee,karasik2011-ntp}, reducing which lowers
the minimum detectable energy and reduces the thermal
response time.  These parameters can be adjusted to optimize the detector
performance, within the constraints of the readout method.

In this work, we propose a mesoscopic superconducting
bolometer/calorimeter for single-photon and continuous power detection
in the GHz--THz frequency range. The operation is based on
temperature-dependent kinetic inductance change of a superconducting
weak link, resulting to temperature--to--phase conversion (TPC) in a
superconducting ring, and its nonlocal readout via an integrated
SQUIPT sensor (see Fig.~\ref{fig:setup}).
The measurement scheme can accommodate a small detector volume, enabling
small heat conductivity and heat capacity, which results in reduced intrinsic thermal noise
and low minimum detectable energy and response time.
In the weak-link device, these quantities are also tunable via the magnetic flux.
For the readout scheme, we predict temperature
sensitivities of tens of $\unit{nK/\sqrt{Hz}}$ in a temperature range tunable
over $T=\unit[10]{mK}\ldots\unit[1]{K}$ with the choice of the
magnetic flux.
We provide an appropriate simplified theoretical model
for the operation and thermal response, and analyze the main
performance characteristics in the bolometric and calorimetric modes.
In calorimetric operation, we predict signal-to-noise ratios up to
$100$ in the $\unit[10]{GHz}\ldots\unit[10]{THz}$ range, and a
resolving power larger than $100$ for single photons at detector bath
temperatures of $\unit[50]{mK}$, with characteristic thermal time
$\tau\sim\unit[10\ldots500]{\mu{}s}$. In the bolometric mode, noise
equivalent power (NEP) down to
$\text{NEP}\approx\unit[10^{-22}]{W/\sqrt{Hz}}$ at
$T_{bath}\sim\unit[50]{mK}$ is predicted, and mainly limited by the
thermal fluctuation noise in the detector.

The structure of the manuscript is as follows.  In
Sec.~\ref{sec:detector} we outline the operation principle and the
theoretical background for the temperature--to--phase conversion and
its SQUIPT detection. In Sec.~\ref{sec:nanocalorimeter} we discuss the
thermal model for the system, and the main performance characteristics
of the calorimetric operation mode.  Section~\ref{sec:nanobolometer}
discusses the performance characteristics in the bolometric mode, and
Sec.~\ref{sec:conclusions} concludes with discussion.

\section{Detector}
\label{sec:detector}
\subsection{Operating principle}

We consider a dc superconducting quantum interference device (SQUID)
based on two superconductor-normal metal-superconductor (SNS)
Josephson junctions [see Fig.~\ref{fig:setup}(a)]. Each junction consists
of two superconducting leads coupled to a N wire of length $L$ through
highly \emph{transmissive} interfaces.  The contact with S induces
superconducting correlations in the N regions through \emph{proximity}
effect which is responsible for the supercurrent flow through the
junctions as well as for the modification of the wires densities of
states (DOSs). The transverse dimensions of the wires are assumed to be much
smaller than $L$ so that they can be considered as
quasi-one-dimensional.

Both SNS junctions are assumed to be \emph{short}, i.e., they satisfy
the condition $\Delta_0\lesssim \hbar D/L^2$, where $\Delta_0$ is the
S energy gap, and $D$ is the diffusion constant of N.  The above
choice is dictated by three main reasons: (i) Drastic reduction of the
detector sensing element volume in order to achieve a substantial
improvement of the sensor performance. (ii) The analytical solutions
for both the supercurrent and the quasiparticle density of states
(DOSs) in the weak-links are well known in the short-junction limit,
which somewhat simplifies the whole analysis of the detector. (iii)
For short junctions, the phase-dependent response of the DOSs in the
read-out junction is maximized leading to enhanced readout transduction
sensitivity.

The variables $\varphi_d$ and $\varphi_r$ denote the
macroscopic quantum phase differences across the detector and readout
junction, respectively.  By neglecting the ring kinetic inductance it
follows from fluxoid quantization that
\begin{equation}
\varphi_d+\varphi_r=2\pi\Phi_{ext}/\Phi_0+2k\pi, 
\label{fluxoid}
\end{equation}
where $\Phi_{ext}$ is the applied magnetic flux, $k$ is an
integer, and $\Phi_0\simeq 2.067\times 10^{-15}$ Wb is the flux
quantum.

We suppose the S loop to be backed at a distance $L_S$ from the NS
interfaces of the detector junction by \emph{Andreev mirrors}, \cite{nahum1993-uhe}
superconductors S$_1$ with a large energy gap $\Delta_1\gg \Delta_0$
in good electric contact, so that S$_1$ blocks out-diffusion of the
heat absorbed by the N wire and the S electrodes [see
  Fig.~\ref{fig:setup}(b)], as quasiparticle energy cannot escape through
the S/S$_1$ interfaces due to Andreev reflections.  The above design
for the detector region allows thermal relaxation to
occur predominantly via the slower electron-phonon mechanism in the N
and S regions. In order to retain the larger kinetic inductance of the
short junction link determined by $\Delta_0$ instead of $\Delta_1$, we
choose below the length $L_S\gg\xi_0$ where $\xi_0=\sqrt{\hbar
  D/\Delta_0}$ is the coherence length of the weaker superconductor.

The external incident radiation is coupled to the N region via a
suitable antenna. If it is coupled as in Fig.~\ref{fig:setup}, the
absorbed power is distributed both to the detector and the readout
junctions, and divided evenly if the junctions are
identical. Different designs may also be able to concentrate the power
dissipation mainly in the detector junction, between the Andreev
mirrors. Here, we neglect the impedance of the ring itself, as it is
small in the GHz--THz range compared to the junctions. In the
discussion that follows, we ignore heat losses due to part of the
power being dissipated in the readout junction.

The antenna impedance matching is influenced by the impedance $Z_d$ of
the short detector junction, which is qualitatively similar to that of
superconductors. \cite{kos2013-fda,abrikosov1959-shf} At low
temperatures and in the short-junction limit, the dissipative component has a frequency threshold
determined by the DOS gap $\varepsilon_g=\Delta_0|\cos\frac{\varphi_d}{2}|$ in
the junction:
$\Re{}Z_d^{-1}(\omega)\sim{}(R_N^{d})^{-1}\theta(\hbar\omega-2\varepsilon_g)+(R_N^{d})^{-1}e^{-\varepsilon_g/k_BT}$.
At frequencies above the gap, the inductive component is of order
$\Im{}Z_d^{-1}(\omega)\sim{}(R_N^{d})^{-1}\pi\Delta_0/(\hbar\omega)$. As a
consequence, we expect that good radiation coupling can be achieved at
frequencies $\nu>2\varepsilon_g/h$ by matching the normal-state resistance $R_N^{d}$
of the junction to the antenna. At frequencies below the DOS gap, the
ability of the proximized normal wire to absorb energy in linear
response becomes suppressed at low $T$ --- which is a generic
limitation of superconducting absorber elements. For large enough
excitation amplitude --- ie. phase oscillation induced by incoming
radiation pulse being $\delta\phi_d(t)=eV/(\hbar\omega)\gtrsim1$ where
$V$ is the voltage amplitude across the junction --- multiphoton
events can contribute to the absorption also at lower
frequencies.

It is beneficial for the performance of the device if the readout
junction remains at a low temperature compared to the detector
junction, even if a part of the input power heats it.  This condition is enforced by the presence of the
superconducting tunnel probe and the lack of Andreev mirrors in the
readout junction, resulting to electronic heat out-diffusion that is
larger by a factor of $\sim{}e^{(\Delta_1-\Delta_0)/k_BT}\gg1$.
Depending on the parameters, it is possible to supplement this by an
additional tunnel-coupled N cooling fin \cite{martinez2015-reh} in the
readout junction, but as we discuss in
Sec.~\ref{sec:nanobolometer-thermal}, this is likely not necessary
for the parameters we consider.

\subsection{Model}
\label{eq:phasemodel}

In the short-junction limit, the Josephson current ($I_c^{d,r}$)
flowing through the detector and readout weak-links at temperature $T$
can be written as
\cite{kulik1975-cmt,heikkila2002-sdo}
\begin{eqnarray}
  I_c^{d,r}(T,\varphi_{d,r})=\frac{\pi\Delta(T)}{eR_N^{_{d,r}}}\text{cos}\left(\frac{\varphi_{d,r}}{2}\right) \int_{\Delta(T)\cos (\varphi_{d,r}/2)}^{\Delta(T)}d\varepsilon
  \nonumber\\
  \times \frac{1}{\sqrt{\varepsilon^2-\Delta^2(T)\text{cos}^2\left(\frac{\varphi_{d,r}}{2}\right)}} \tanh\left(\frac{\varepsilon}{2 k_B T}\right),\,\,\,\,\,\,\,
  \label{critcurrent}
\end{eqnarray}
where $\Delta (T)$ is the BCS temperature-dependent pairing potential
in S, $R_N^{d,r}$ is the normal-state resistance of detector (readout)
junction, $k_B$ is the Boltzmann constant, and $e$ is the electron
charge.  In the limit of zero temperature ($T=0$),
Eq.~\eqref{critcurrent} reduces to \cite{kulik1975-cmt}
\begin{equation}
I_c^{d,r}(0,\varphi_{d,r})=\frac{\pi\Delta(0)}{eR_N^{_{d,r}}}\text{cos}\left(\frac{\varphi_{d,r}}{2}\right)\artanh\left[\sin \left(\frac{\varphi_{d,r}}{2}\right)\right].
\label{critcurrentT0}
\end{equation}
In Eq.~\eqref{critcurrent}, $R_N^{d,r}=\rho L/A^{d,r}$,
$\rho=(\nu_Fe^2D)^{-1}$ is the wire resistivity, $\nu_F$ the density
of states at the Fermi level in N, and $A^{d,r}$ is the wire cross
section of the detector (readout) Josephson junction.  Moreover, in
the following the electron temperature in the detector junction will
be denoted with $T_e$ ($T\equiv T_e$) whereas that in the read-out
weak-link is supposed to coincide with the lattice temperature,
$T\equiv T_{bath}$ (see discussion in
Sec.~\ref{sec:nanobolometer-thermal}).

\begin{figure}
  \includegraphics[width=\columnwidth]{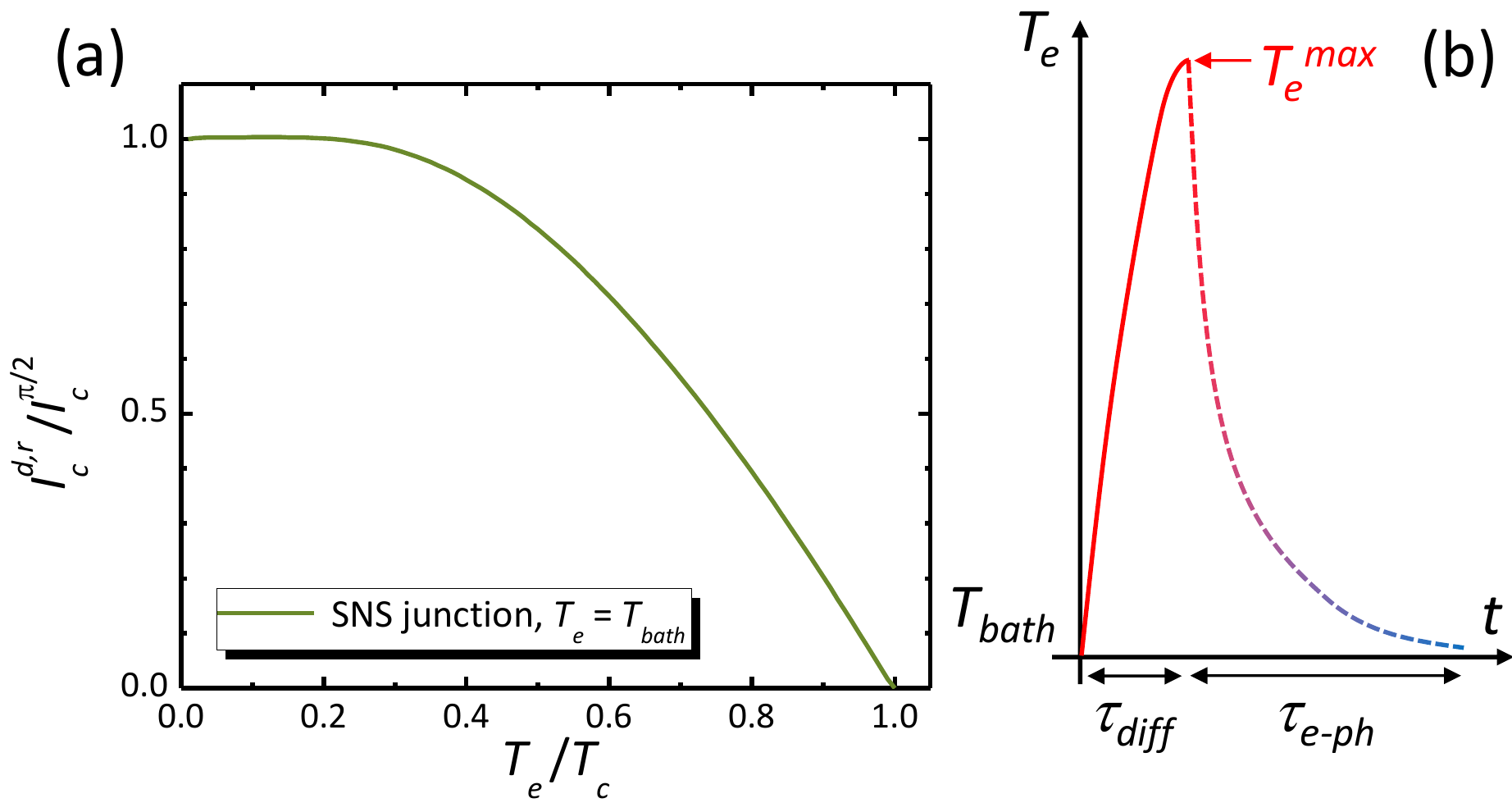}\vspace{-3mm}
  \caption{(a) Josephson current $I^{d,r}$ vs
    $T_e=T_{bath}$ calculated at $\varphi_{d,r}=\pi/2$. $T_{bath}$ and
    $T_c$ are the bath and critical temperature of the S loop,
    respectively, whereas $I_c^{\pi/2}$ is the zero-temperature
    Josephson current at $\varphi_{d,r}=\pi/2$.  (b) Scheme of time
    ($t$) evolution of the electronic temperature $T_e$ in the
    detector junction after absorption of a photon at energy $h\nu$.
    $T_e^{max}$ is the temperature reached in the weak-link after the
    arrival of a single photonic event.  $\tau_{diff}$ represents the
    diffusion time across the N wire whereas $\tau_{e-ph}$ is the
    electron-phonon relaxation time. In the present setup
    $\tau_{diff}\ll \tau_{e-ph}$ due to the short length of the N
    wire, and the low temperature at which the detector is conceived
    to operate.  }
  \label{fig:critcur}
\end{figure}

The electromagnetic energy $h\nu$ absorbed in the detector
junction elevates the temperature $T_e$ in N and in the lateral
portion of S thereby leading to a decrease of the dissipationless
supercurrent $I_c^{d,r}$ circulating (for $\Phi_{ext}\neq 0$) in the
superconducting loop.  The temperature dependence of the junctions
critical current is shown in panel (a) of Fig.~\ref{fig:critcur}, being
evaluated from Eq.~\eqref{critcurrent} for $\varphi_d=\varphi_r=\pi/2$
by setting $T\equiv T_e=T_{bath}$.  Note that the critical
supercurrent is almost saturated for $T_e\lesssim 0.25T_c$ ($T_c$ is
the critical temperature of S), and decreases linearly with
temperature around $T_e\approx T_c$.

Such a temperature-induced suppression of $I_c^{d,r}$ yields a finite
variation of the phase drop across both SNS junctions owing to the
following reasons: i) Conservation of the supercurrent circulating
along the S loop, and ii) fluxoid quantization in the interferometer
[Eq.~\eqref{fluxoid}].  As a consequence, for a given $\Phi_{ext}$,
the phases $\varphi_d$ and $\varphi_r$ can be determined for any
$T_{e}$ and $T_{bath}$ from condition i), i.e., by solving the
equation
\begin{equation}
I_c^{d}(T_{e},\varphi_{d})=I_c^{r}(T_{bath},\varphi_{r}). 
\label{phase}
\end{equation}
This is at the origin of the \emph{temperature-to-phase} conversion
(TPC) process.  By defining the parameter
$\alpha=R_N^d/R_N^r=A^r/A^d$ as the degree of asymmetry of the SQUID
junctions, we can solve Eq.~\eqref{phase} for the phase $\varphi_r$
existing across the readout junction in order to investigate in detail
the full TPC process. We note that the asymmetry parameter can also be
written as $\alpha=L_J^d/L_J^r$ when both junctions are at the same temperature,
where $L_J^{d,r}$ is the kinetic
inductance of the detector (readout) junction, so that it immediately
expresses which junction of the interferometer will mainly determine
the phase biasing of the superconducting loop for a given external
flux $\Phi_{ext}$. In particular, for $\alpha\gg 1$ the phase drop
along the ring will occur predominantly across the \emph{detector}
junction whereas in the opposite situation ($\alpha\ll 1$) the phase
drop will occur mainly across the \emph{readout} weak-link.

\begin{figure}
  \includegraphics[width=\columnwidth]{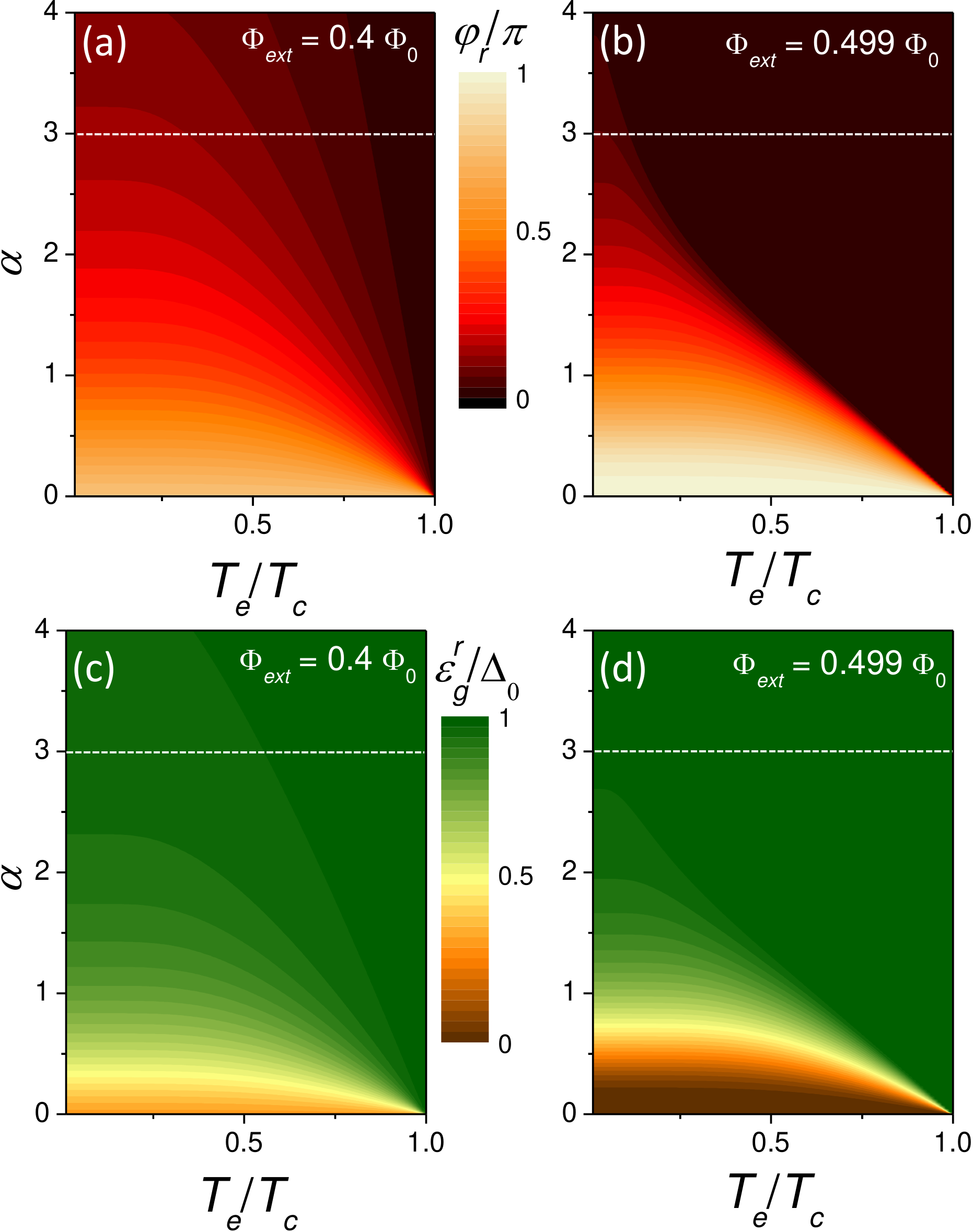}\vspace{-3mm}
  \caption{ Top panels: Contour plot of the phase $\varphi_r$ across
    the readout junction vs $\alpha$ and $T_e$ calculated for
    $\Phi_{ext}=0.4\Phi_0$ (a) and $\Phi_{ext}=0.499\Phi_0$ (b) at
    $T_{bath}=0.01 T_c$.  Bottom panels: Contour plot of the minigap
    $\varepsilon_g^r$ in the N region of the readout junction vs
    $\alpha$ and $T_e$ calculated for $\Phi_{ext}=0.4\Phi_0$ (c) and
    $\Phi_{ext}=0.499\Phi_0$ (d) at $T_{bath}=0.01 T_c$.  Dashed lines
    in all panels positioned at $\alpha=3$
    correspond to the value chosen for the interferometer
    in all forthcoming calculations, unless differently stated.
    $\Delta_0$ denotes the zero-temperature energy gap in S
    corresponding to the critical temperature $T_c \simeq
    \Delta_0/1.764k_B$.  }
  \label{fig:phase}
\end{figure}

Top panels of Fig.~\ref{fig:phase} show the phase $\varphi_r$ calculated
from Eq.~\eqref{phase} as a function of $T_e$ and $\alpha$ for
$\Phi_{ext}=0.4\Phi_0$ [(a)] and $\Phi_{ext}=0.499\Phi_0$ [(b)] at
$T_{bath}=0.01 T_c$. It turns out that the largest TPC effect occurs
by setting $\Phi_{ext}$ around $\sim0.5\Phi_0$ where the phase
response can be very sharp [see panel (b)].  In particular, by
choosing values for $\alpha$ around or slightly smaller than $1$
yields sizable phase modulation amplitudes in the readout junction at higher temperatures
$T_e$. By contrast, for $\alpha >1$, the modulation amplitude occurs
for smaller temperatures, although slightly reduced.  

The parameter $\alpha$ also controls the phase difference over the
detector junction $\varphi_d$ and therefore the energy gap
$\varepsilon_g^d=\Delta_0\cos(\frac{\varphi_d}{2})$. As discussed above, this
provides a frequency threshold $2\varepsilon_g^d$ below which it can be
difficult to impedance match the detector junction to an antenna. To
reduce $\varepsilon_g^d$, a larger $\alpha$ should be chosen --- which however
reduces the sensitivity of TPC. Below, we choose this tradeoff at
$\alpha=3$ (dashed lines in Fig.~\ref{fig:phase}), corresponding
to $2\varepsilon_g/h=\unit[10]{GHz}$.
If a higher frequency threshold is acceptable, better performance
characteristics can be achieved by values closer to $\alpha=1$.

\subsection{Readout weak-link behavior}

The temperature-induced suppression of the critical current may
thereby yield a variation of phase drop ($\delta \varphi_r$) across
the readout junction.  By measuring $\delta \varphi_r$ with a suitable
setup would enable to assess with accuracy the electronic temperature
$T_e$, and hence the radiation absorbed by the SNS weak-link of the
detector.  Yet, owing to \emph{proximity effect}, $\varphi_r$
affect as well the spectral characteristics of the corresponding N region, for
instance, by determining the exact shape of the local quasiparticle
density of states (DOSs) in the N wire.  In the following we
show that by probing the phase-induced variations of the DOSs with a
superconducting quantum interference proximity transistor (SQUIPT)
implemented in the readout junction is a simple and effective way to
get direct and detailed information about $T_e$-driven phase changes.

%%%%%%%%%%%%%%%%%%%%%%%%%%%%%%%%%%%%%%%%%%%%%%% - SQUIPT behavior - %%%%%%%%%%%%%%%%%%%%%%%%%%%%%%%%%%%%%%%%%%%%%%%%%%%%%%%%%%%%%%%%%%%%%

The SQUIPT [shown on the upper part of the scheme of
  Fig.~\ref{fig:setup}(a)] consists of a superconducting tunnel junction
with normal-state resistance $R_p$ coupled to the middle of the N wire
of the readout junction \cite{giazotto2010-sqi}.  So far, SQUIPTs have
been implemented in a few geometrical configurations
\cite{giazotto2010-sqi,strambini2016-ost,meschke2011-tsp} and with different materials combinations
\cite{dambrosio2015-nmt,giazotto2010-sqi,ronzani2016-pdc}.  For the sake of clarity, we assume
here the probing electrode (of width $w$) to be made of the same
superconducting material S as the SQUID ring.  $\varphi_r$ changes
induced by temperature $T_e$ variations in the detector junction
affect the readout N wire DOSs, and thereby the current $I_p$ vs
voltage characteristic of the superconducting tunnel junction biased
at voltage $V$ [see Fig.~\ref{fig:setup}(a)].

Let us now analyze the N wires DOSs ($\mathcal{N}_N^{d,r}$) in the SNS
junctions.  In the short-junction limit, which is the relevant one for
the present case, we have the well-know result
(cf. e.g. Refs.~\onlinecite{heikkila2002-sdo,artemenko1979-ton}):
\begin{equation}
  \begin{split}
  \mathcal{N}_N^{d,r}(x,\varphi_{d,r},\varepsilon,T)
  =
  \Re\sqrt{\frac{(\varepsilon+i\Gamma)^2}{(\varepsilon+i\Gamma)^2-\Delta^2(T)\cos^2\frac{\varphi_{d,r}}{2}}}
  \\
  \times
  \cosh\Bigl(
  \frac{2x-L}{L}\arcosh\sqrt{\frac{(\varepsilon+i\Gamma)^2-\Delta^2(T)\text{cos}^2\frac{\varphi_{d,r}}{2}}{(\varepsilon+i\Gamma)^2-\Delta^2(T)}}
  \Bigr),
  \end{split}
\end{equation}
where $\varepsilon$ is the energy relative to
the chemical potential of the superconductors, $x \in[0,L]$ is the
spatial coordinate along the N wires, and $\Gamma$ is the Dynes
parameter, accounting for broadening in S.

$\mathcal{N}_N^{d,r}$ exhibits a minigap
$\varepsilon_g^{d,r}(\varphi_{d,r,}T)=\Delta(T)|\text{cos}(\varphi_{d,r}/2)|$
for $|\varepsilon|\leq \varepsilon_g^{d,r}$ whose amplitude depends on
$\varphi_{d,r,}$, and is spatially constant along the N wires. In
particular, $\varepsilon_g^{d,r}=\Delta$ for $\varphi_{d,r,}=0$ and
decreases by increasing the value of the phase, vanishing at $\pi$.
Therefore, the quasiparticle spectrum in the N region can vary from
that of a \emph{gapped} superconducting material (for
$\varphi_{d,r,}=0$) to that of a \emph{gapless} normal conductor (at
$\varphi_{d,r,}=\pi$) just by changing the phase across the weak-link.

The impact of the electronic temperature $T_e$ of the detector on the
DOSs in the readout N weak-link is displayed in the two bottom panels
of Fig.~\ref{fig:phase} where the minigap amplitude $\varepsilon_g^{r}$ is
plotted as a function of $T_e$ and $\alpha$ for $\Phi_{ext}=0.4\Phi_0$
[(c)] and $\Phi_{ext}=0.499\Phi_0$ [(d)], both calculated at
$T_{bath}=0.01 T_c$.  These results show that the minigap behavior is
qualitatively similar to that of $\varphi_r$, and confirm that for
$\alpha=3$ it is possible to obtain sufficient $T_e$-induced
modulation of $\varepsilon_g^{r}$. The above value for $\alpha$ will
be set in all forthcoming calculations, unless differently stated, to
evaluate the response and performance of the nanodetector.

%%%%%%%%%%%%%%%%%%%%%%%%%%%%%%%%%%%%%% - SQUIPT current - %%%%%%%%%%%%%%%%%%%%%%%%%%%%%%%%%%%%%%%%%%%%%%%%%%%%%%%%%%
\subsection{SQUIPT response}

Let us now turn on discussing the behavior of the SQUIPT current vs
voltage characteristics which allow to understand how to exactly
operate the superconducting interferometer for radiation detection in
the present setup.  The current flowing through the superconducting
tunnel probe of the SQUIPT in the readout junction is dominated by
quasiparticles, and can be written as
\cite{giazotto2011-hsq}
\begin{equation}
I_p(V)=\frac{1}{ewR_p}\int_{\frac{L-w}{2}}^{\frac{L+w}{2}}dx\int_{-\infty}^{\infty}d\varepsilon \mathcal{N}_N^r(x,\varepsilon,\varphi_r)\mathcal{N}_{S}^p(\tilde{\varepsilon})F(\varepsilon,\tilde{\varepsilon}),
\label{currentSQUIPT}
\end{equation}
where
$\mathcal{N}_S^p(\varepsilon,T_e)=|\Re[(\varepsilon+i\Gamma)/\sqrt{(\varepsilon+i\Gamma)^2-\Delta^2(T_e)}]|$ is the BCS normalized
DOS of the S probe electrode at temperature $T_{bath}$,
$\tilde{\varepsilon}=\varepsilon-eV$,
$F(\varepsilon,\tilde{\varepsilon})=[f_0(\tilde{\varepsilon})-f_0(\varepsilon)]$,
$f_0(\varepsilon)$ is the Fermi-Dirac energy distribution function,
and $R_p$ is the normal-state tunneling resistance of the probing
junction.  In the following calculations we assume for simplicity that
the superconductor forming the probing electrode is identical to that
realizing the loop so that
$\mathcal{N}_{S}^p(\varepsilon,T)=\mathcal{N}_{S}^d(\varepsilon,T)$. Moreover,
we set $w=L/3$ and $R_p=10^5\Omega$ as characteristic parameters of
the SQUIPT readout.  For calorimetric operation, we also assume
measurement of the current is possible on bandwidths comparable to the
relevant inverse thermal relaxation time of the detector junction (see
below) --- possible fast readout schemes are discussed in
Refs.~\onlinecite{gasparinetti2015-fet,saira2016-dtj}.

\begin{figure}
  \includegraphics[width=\columnwidth]{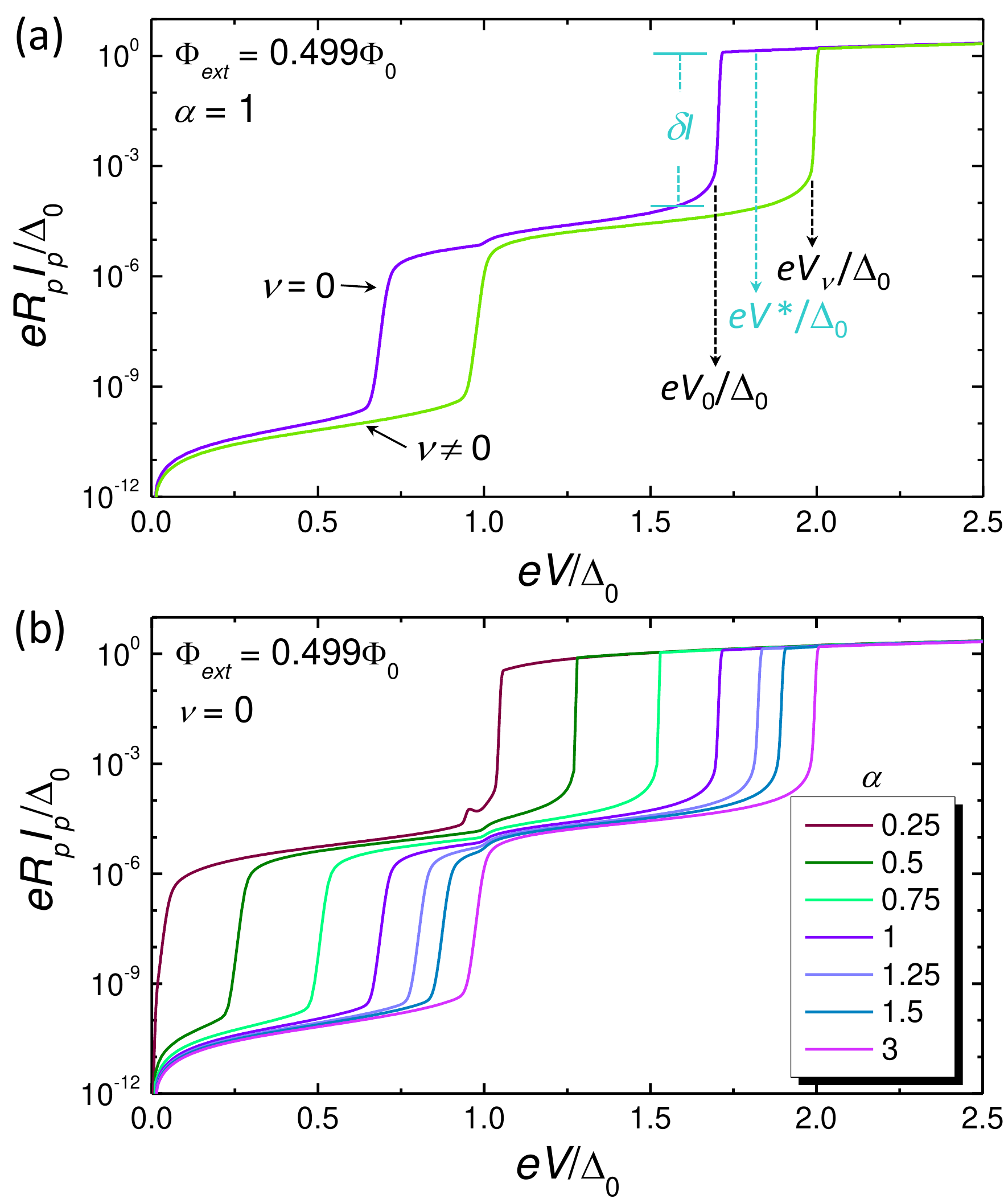}\vspace{-1mm}
  \caption{(a) Current ($I_p$) vs voltage ($V$) characteristics of the
    SQUIPT readout junction in the absence ($T_e=T_{\mathrm{bath}}$, $\nu=0$) and for finite
    incoming radiation ($T_e>T_{\mathrm{bath}}$, $\nu\neq0$) calculated for
    $\Phi_{ext}=0.499\Phi_0$, $\alpha=1$ and $T_{bath}=0.01T_c$. The
    voltage onset of large quasiparticle tunneling for zero ($h\nu=0$) and finite
    frequency ($h\nu\ne0$) are indicated (black-dashed arrows). The bias voltage of
    the SQUIPT readout junction is denoted with $V^{\ast}$ whereas
    $\delta I$ indicates the current variation due to the absorption of
    a photon of energy $h\nu$.  (b) $I_p$vs $V$ calculated for
    $\Phi_{ext}=0.499\Phi_0$, $T_{bath}=0.01T_c$, $\nu=0$ and for
    several values of $\alpha$. The degree of asymmetry of the
    interferometer determines the onset of large quasiparticle
    tunneling in the absence of radiation.
  }
  \label{IV}
\end{figure}

Figure \ref{IV}(a) illustrates the general behavior of the low-temperature SQUIPT
current vs voltage characteristics $I_p(V)$ at
$\Phi_{ext}=0.499\Phi_0$ in the absence of radiation (i.e., $h\nu=0$, $T_e=T_{\mathrm{bath}}$),
and for a nonzero radiation ($h\nu\neq 0$, $T_e>T_{\mathrm{bath}}$) heating the
detector junction.  In particular, the onset of large quasiparticle
tunneling occurs at an energy corresponding to the sum of the gaps in
the superconducting probe and the N proximity layer, as expected for a
tunneling process through an SIS' tunnel junction \cite{tinkham1996-its}.
Therefore, in the absence of radiation this onset appears at
$eV_0=\Delta(T)+\varepsilon_g^r(T,\nu=0)$ whereas under the effect of
radiation of energy $h\nu$ the onset occurs at
$eV_{\nu}=\Delta(T)+\varepsilon_g^r(T,\nu)$.  Moreover, $2\Delta(T)/e\ge{}V_{\nu}>V_0$
since $\Delta(T)\ge\varepsilon_g^r(T,\nu)>\varepsilon_g^r(T,\nu= 0)$ as a
consequence of heating in the detector junction originating from
absorption of radiation [see Figs. \ref{fig:phase}(c,d)].  From this it
follows that, by biasing the SQUIPT at voltage $V^{\ast}$ (with
$V_0\lesssim V^{\ast}\lesssim V_{\nu}$), the absorption of a photon
will yield a reduction ($\delta I$) of the current $I_p$ flowing
through the tunneling probe. As a consequence, a direct radiation
readout can be performed with the SQUIPT by a simple measurement of
its current at fixed bias voltage.

The impact of asymmetry of the two SNS Josephson junctions forming the
SQUID on the SQUIPT characteristics is displayed in Fig. \ref{IV}(b)
which shows $I_p$ vs $V$ calculated for selected values of $\alpha$ at
$\Phi_{ext}=0.499\Phi_0$, and $\nu=0$. The figure shows that increasing
$\alpha$ leads to $V_0$ approaching $2\Delta_0$, as expected from the
increase of $\varepsilon_g^r$ in the readout junction [see
  Figs.~\ref{fig:phase}(c,d)]. Therefore, a precise tuning of the SQUIPT
working voltage $V^{\ast}$ can, in principle, be achieved by setting
the asymmetry of the junctions forming the SQUID.

For the value $\alpha=3$ used in the remainder of the text,
at $T_e=T_{bath}=0.01\Delta_0$ and $\Phi_{ext}=0.499\Phi_0$,
$\varphi_r\approx0.07\pi$, and
the voltage threshold for $\nu=0$ [see Fig.~\ref{IV}(a)] resides at
$eV_0\approx{}1.994\Delta_0$.

\subsection{Temperature-to-current conversion and noise analysis}

\label{sec:tsensitivity}

Hereafter, we describe the behavior of the SQUIPT corresponding to a
\emph{temperature-to-current} transducer, i.e., the ability of the
system to convert a temperature variation $\delta T_e$ in the detector
weak link into a current change $\delta I$ in the readout tunnel junction.

\begin{figure}
  \includegraphics[width=\columnwidth]{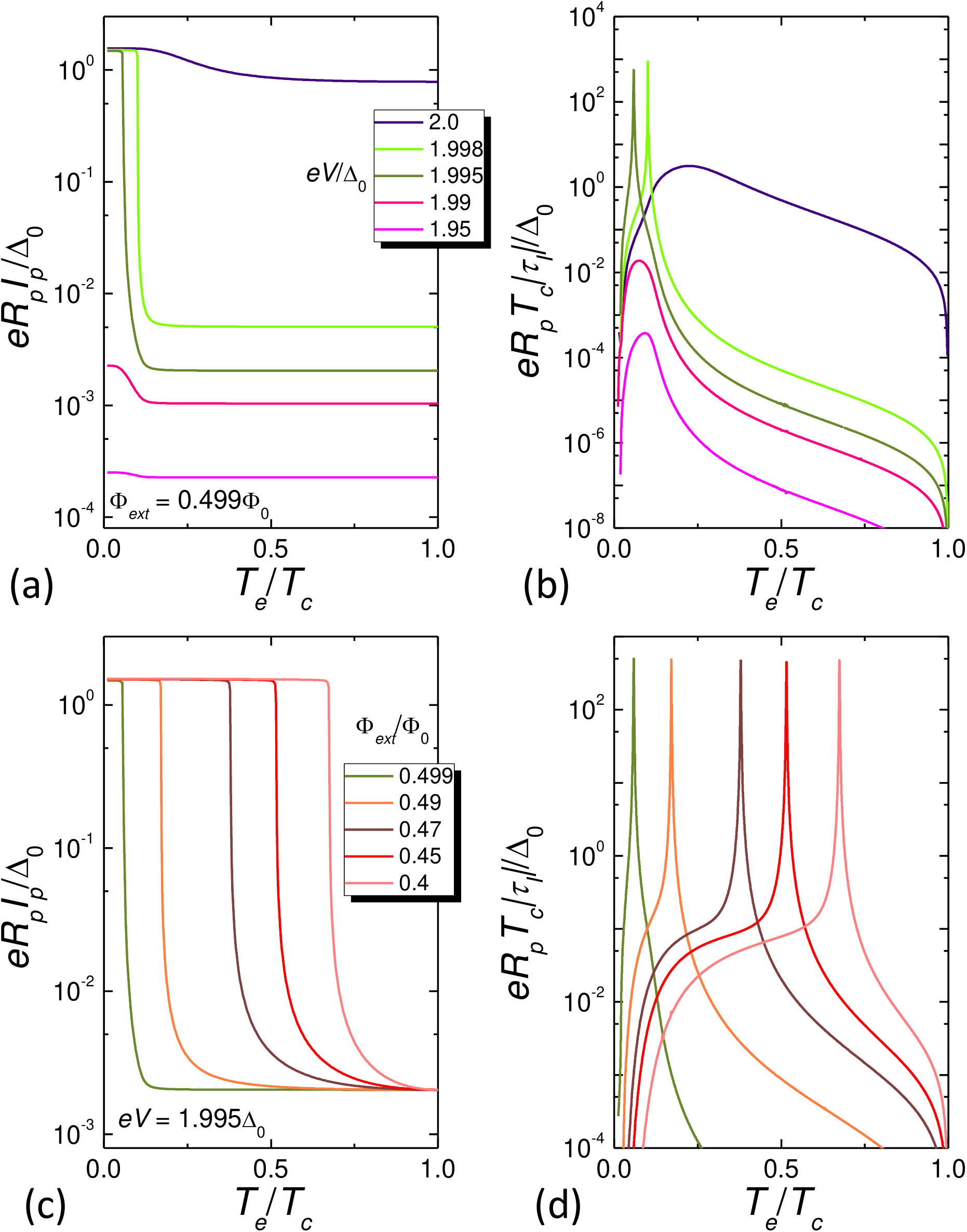}\vspace{-1mm}
  \caption{(a) Current $I_p$ vs temperature $T_e$ characteristics of
    the SQUIPT readout junction calculated for a few values of $V$ at
    $\Phi_{ext}=0.499\Phi_0$.  (b) Absolute value of the
    temperature-to-current transfer function $|\tau_I|$ vs $T_e$
    calculated for the same parameters as in panel (a).  (c) $I_p$ vs
    $T_e$ characteristics calculated for a few values of $\Phi_{ext}$
    at $V=1.995\Delta_0/e$.  (d) Absolute value of the
    temperature-to-current transfer function $|\tau_I|$ vs $T_e$
    calculated for the same parameters as in panel (c).  In all these
    calculations we set $\alpha=3$, and $T_{bath}=0.01T_c$.
  }
  \label{resp}
\end{figure}

Figure \ref{resp}(a) shows the dependence of the current $I_p$ through
the probing junction on the temperature $T_e$ for different values of
$V$ by keeping fixed the external flux $\Phi_{ext}=0.499\Phi_0$, for
$\alpha=3$ and $T_{bath}=0.01T_c$. In particular, the current turns
out to be strongly dependent on $T_e$ for specific values of the
biasing voltage, in particular, we note that for
$V=V^{\ast}=1.995\Delta_0/e$ and $V=V^{\ast}=1.998\Delta_0/e$, $I_p$ can
vary significantly for temperatures $T_e$ in the range $\sim 0.05 T_c\ldots
0.2 T_c$.  The above $V^{\ast}$ values stem from the chosen parameters
of the structure, and will lead to a high sensitivity for radiation
detection.  By contrast, for other bias voltage values the current
response is somewhat moderate in the whole range of temperatures.

A figure of merit which is useful to characterize the readout
weak-link performance is represented by the temperature-to-current
transfer function, $\tau_I=\partial I_p/\partial T_e$, which is shown
in Fig. \ref{resp}(b) for the same parameters as in panel (a). For
$V=V^{\ast}=1.995\Delta_0/e$ and $V=V^{\ast}=1.998\Delta_0/e$ $\tau_I$
obtains the largest values in the range $\sim 0.05T_c\ldots 0.2 T_c$,
which is expected from the corresponding behavior of the current in
the same temperature window.

Figure \ref{resp}(c) displays the dependence of $I_p$ on $T_e$ for
different values of the applied magnetic flux $\Phi_{ext}$ by keeping
fixed the bias voltage $V=1.995\Delta_0/e$, for $\alpha=3$ and
$T_{bath}=0.01T_c$. Approaching the $0.5\Phi_0$ flux leads
to a stronger response of tunneling current through the readout
junction. On the other side, values of $\Phi_{ext}$ far away from
$0.5\Phi_0$ correspond in general to a weaker dependence of $I_p$ on the
temperature. The corresponding transfer functions are shown in
Fig. \ref{resp}(d), and are calculated for the same parameters as in
panel (c).

The intrinsic temperature sensitivity (temperature \emph{noise}) per
unit bandwidth of the probe junction ($s_T$) is related to its
current-noise spectral density ($\mathcal{S}_I$) as
\begin {equation}
  s_T=\frac{\sqrt{\mathcal{S}_I}}{|\tau_I|},
  \label{temperaturenoise}
\end{equation}
where $\tau_I$ is the temperature-to-current transfer function
discussed before.  The low-frequency current noise spectral density of the
tunnel probe is \cite{rogovin1974-fpt}
\begin{equation}
  \mathcal{S}_{I,tun}(V)
  =2e I_p(V) \coth\Bigl(\frac{eV}{2k_BT_{bath}}\Bigr).
  \label{shot}
\end{equation}
We note that Eq.~\eqref{shot} describes both regimes of shot noise,
i.e., $k_BT_{bath}\ll eV$, and thermal noise, i.e., $eV\ll
k_BT_{bath}$, and holds in the tunneling limit.

\begin{figure}
  \includegraphics[width=\columnwidth]{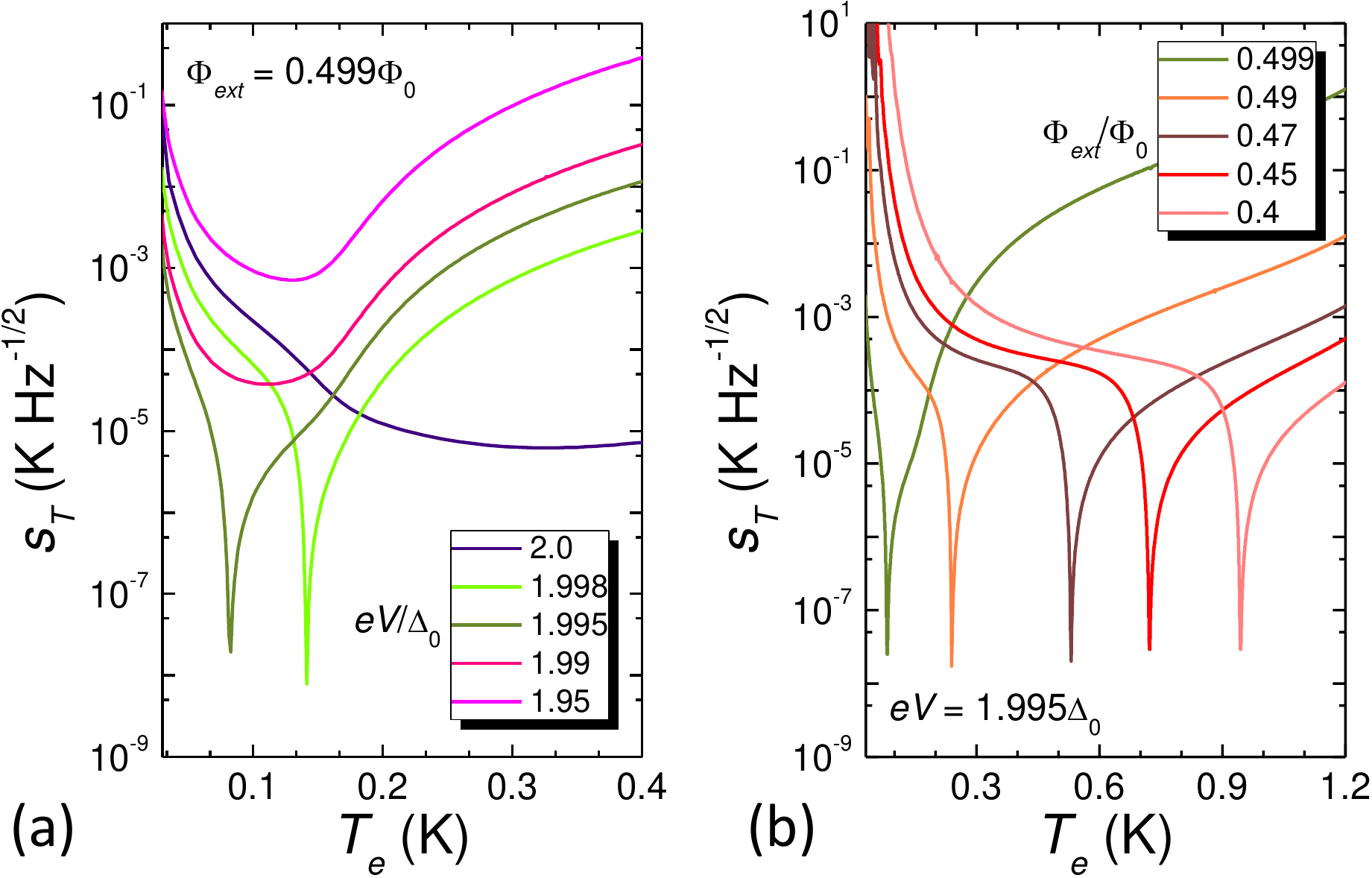}\vspace{-1mm}
  \caption{(a) Temperature sensitivity $s_T$ vs temperature $T_e$
    calculated for a few values of $V$ at $\Phi_{ext}=0.499\Phi_0$.
    (b) $s_T$ vs $T_e$ calculated for a few values of $\Phi_{ext}$ at
    $V=1.995\Delta_0/e$. In all the calculations we set $\alpha=3$, and
    $T_{bath}=0.01T_c$.
  }
  \label{tnoise}
\end{figure}

In Fig. \ref{tnoise} (a) we show the temperature noise $s_T$ vs $T_e$
calculated for selected values of $V$ at $\Phi=0.499\Phi_0$. The
maximum temperature sensitivity is obtained for $V\sim 1.95\cdots
2.0\Delta_0/e$ where the temperature noise can be of order
$\unit[10^{-8}]{K/\sqrt{Hz}}$. Figure \ref{tnoise} (b) displays the temperature noise
vs $T_e$ calculated for a few values of the external flux at
$V=1.995\Delta_0/e$. In particular, $s_T$ is minimized at different
temperatures depending on the specific value of $\Phi_{ext}$, and obtains
values as low as a few tens of $\unit{nK/\sqrt{Hz}}$, at temperature range tunable
over $T=\unit[10]{mK}\ldots\unit[1]{K}$ with the choice of $\Phi_{ext}$.

%%%%%%%%%%%%%%%%%%%%%%%%%%%%%%%%%%%%%%%%%%%%%%%%%%%%%%%%%%%%%%%%%% - Secion II. Nanocalorimeter - %%%%%%%%%%%%%%%%%%%%%%%%%%%%%%%%%%%%%%%%%%%%%%
\section{Nanocalorimeter operation}
\label{sec:nanocalorimeter}
\subsection{Thermal model}

The operation principle of single-photon detection (i.e., operation as
a \emph{calorimeter}) based on the TPC effect can be understood
by inspecting the scheme displayed in Fig.~\ref{fig:critcur}(b) which shows a
sketch of time evolution of the electronic temperature $T_e$ in the
detector weak-link after the arrival of a photonic event. In
particular, we assume that depending on the photon energy $h\nu$,
$T_e$ is increased with respect to $T_{bath}$ up to $T_e^{max}$
uniformly along the wire over a time scale set by
fast time scales, e.g. the diffusion time $\tau_{diff}=L^2/D$ (see red full line)
and LC time of the superconducting loop. The diffusion time with the typical
parameters chosen for our SNS junctions, is of the order of
$\sim10^{-12}$ s.  After the initial absorption, the electronic
temperature $T_e$ relaxes
towards the bath temperature over a time scale set by the electron-phonon
relaxation time, $\tau_{e-ph}$.  For instance, in the normal state,
this time is given by $\tau_{e-ph}\approx
\frac{k_B^2\nu_F}{0.34\Sigma}T_{bath}^{-3}$ where $\Sigma$ is the
electron-phonon coupling constant.  By setting for instance
$\Sigma=5\times 10^8$ Wm$^{-3}$K$^{-5}$ typical of silver (Ag) we
obtain $\tau_{e-ph}\in[10^{-4}\cdots 10^{-7}]$ s for
$T_e\in[10^{-1}\cdots 1]$ K so that $\tau_{diff}\ll \tau_{e-ph}$ [see
  dashed line in Fig.~\ref{fig:setup}(c)]. As we shall argue, in our system
thanks to superconducting correlations induced in the N regions,
$\tau_{e-ph}$ can be longer than that in the normal
state.

Consider now an event where the detector junction absorbs energy
$\delta{}U=h\nu$.  The energy $\delta{}U(T_e,T_{\rm bath},\Phi_{ext})$
required for increasing the temperature of the detector junction from
$T_{\rm bath}$ to $T_e$ consists of heat $\delta{}Q^d$ of the
quasiparticles in the detector junction, and work
$\delta{}W=\delta{}W^d+\delta{}W^r$ required for changing the phase
differences over the junctions,
$\varphi_{d,r}^{(0)}\mapsto\varphi_{d,r}^{(1)}$.  The energy required
is bounded by free energy differences
$\delta{}W^{r,d}\ge\delta{}F^{r,d}$. Below, we assume that
$\delta{}W^r=\delta{}F^r$ and that the detector does not exchange heat
with other systems (on the short time scales): in the real device the
energy required can be larger due to some heat $\delta{}Q^r\ge0$
entering the readout junction and exiting to heat sinks.  For a device
where the detector junction is impedance matched to the antenna better
than the readout, we expect $\delta{}Q^r<\delta{}Q^d$ so that most of
the absorbed heat is useful for detection.

In the above model, we have
\begin{gather}
  \label{heateq}
  \delta{}U(T_e,T_{\mathrm{bath}},\Phi_{ext})
  =
  U^d(\varphi_d^{(1)}, T_{e}) - U^d(\varphi_d^{(0)},T_{\rm bath})
  \\\notag
  +
  F^r(\varphi_r^{(1)},T_{\rm bath}) - F^r(\varphi_r^{(0)},T_{\rm bath})
  \,.
\end{gather}
Above, the phases $\varphi_{r,d}^{(1)}=\varphi_{r,d}(T_e,T_{bath},\Phi_{ext})$ and
$\varphi_{r,d}^{(0)}=\varphi_{r,d}(T_{bath},T_{bath},\Phi_{ext})$ are the functions of
temperature and flux discussed in Sec.~\ref{eq:phasemodel}.  The
effective total electronic heat capacity for constant $\Phi_{ext}$ is
then defined as
\begin{equation}
  \begin{split}
  C_{tot}^e(T_e,T_{\rm bath})
  &=
  \frac{d}{dT_e}\delta{}U(T_e,T_{\rm bath},\Phi_{ext})
  \\
  &=
  T_e
  \frac{d}{dT_e}S^d(T_e,\varphi_d^{(1)}(T_e,T_{\rm bath},\Phi_{ext}))
  \,,
  \end{split}
\end{equation}
and it accounts for the readjustment of the phases.  The relation to
the quasiparticle entropy $S^d=-\partial_TF^d$ of the detector follows
from Eq.~\eqref{heateq}, $U^{d,r}=F^{d,r}+TS^{d,r}$, and the current
balance
$I^d-I^r=\partial_{\varphi_d}{}F^d(\varphi_d^{(1)},T_e)-\partial_{\varphi_r}F^r(\varphi^{(1)}_r,T_{bath})=0$.

The entropy for short SNS junctions is to a good approximation
\cite{virtanen2017-qes} given by
\begin{align}
  \label{eq:shortjcn-free-energy}
  S^{d}(\varphi,T)
  \simeq
  (
  \mathcal{V}^{d}_S+\mathcal{V}^{d}_N)\mathcal{S}_{BCS}(T)
  -
  \frac{\hbar}{2e} \int_0^\varphi{}d\varphi' \partial_T I_c^{d}(\varphi',T)
  \,,
\end{align}
where the first term also accounts for the phase-independent heat
capacity of the superconducting S part of the detector. The second term corresponds to
the quasiparticle states bound at the junction; the result automatically satisfies the thermodynamic
relation $\partial_\varphi{}S=-\frac{\hbar}{2e}\partial_TI_c$.  The bulk
superconductor entropy density is
\begin{equation}
  \label{eq:bcs-entropy}
  \mathcal{S}_{BCS}(T)=-2\nu_F^Sk_B\int_{-\infty}^{\infty}d\varepsilon \mathcal{N}_{BCS}(T,\Delta(T))
  f_0(\varepsilon,T)\text{ln}[f_0(\varepsilon,T)]
  \,.
\end{equation}
From the above, we can obtain the relations
$\delta{}U(T_e,T_{\rm bath},\Phi_{ext})$ and
$T_e(\delta{}U,T_{\rm bath},\Phi_{ext})$ for given system parameters.

We now fix parameters for the hybrid detector in Fig.~\ref{fig:setup}b.
For the normal part, we choose a 10-nm-thick, 10-nm-wide Ag wire
with $L=200$ nm (volume $\mathcal{V}_N=2\times\unit[10^{-23}]{m^3}=\unit[200]{nm}\times\unit[10]{nm}\times\unit[10]{nm}$). For the superconducting part (S), we consider aluminum (Al), with
$T_c=1.4$~K, $\Delta_0=1.764 k_B T_c$, and select volume $\mathcal{V}_S=2\times 10^{-21}$
m$^3=2\times\unit[500]{nm}\times\unit[20]{nm}\times{}\unit[100]{nm}$,
corresponding to $L_S=\unit[500]{nm}$.
The densities of states are
$\nu_F=\unit[1.0\times 10^{47}]{J^{-1}m^{-3}}$,
$\nu_F^S=\unit[2.15\times 10^{47}]{J^{-1}m^{-3}}$, and $D=0.01$
m$^2$s$^{-1}$.  This corresponds to normal-state resistance of
$R_N\simeq\unit[80]{\Omega}$, but other material choices providing
better impedance matching are also possible. 

\subsection{Temperature response of the detector weak-link}

\begin{figure}
  \includegraphics[width=\columnwidth]{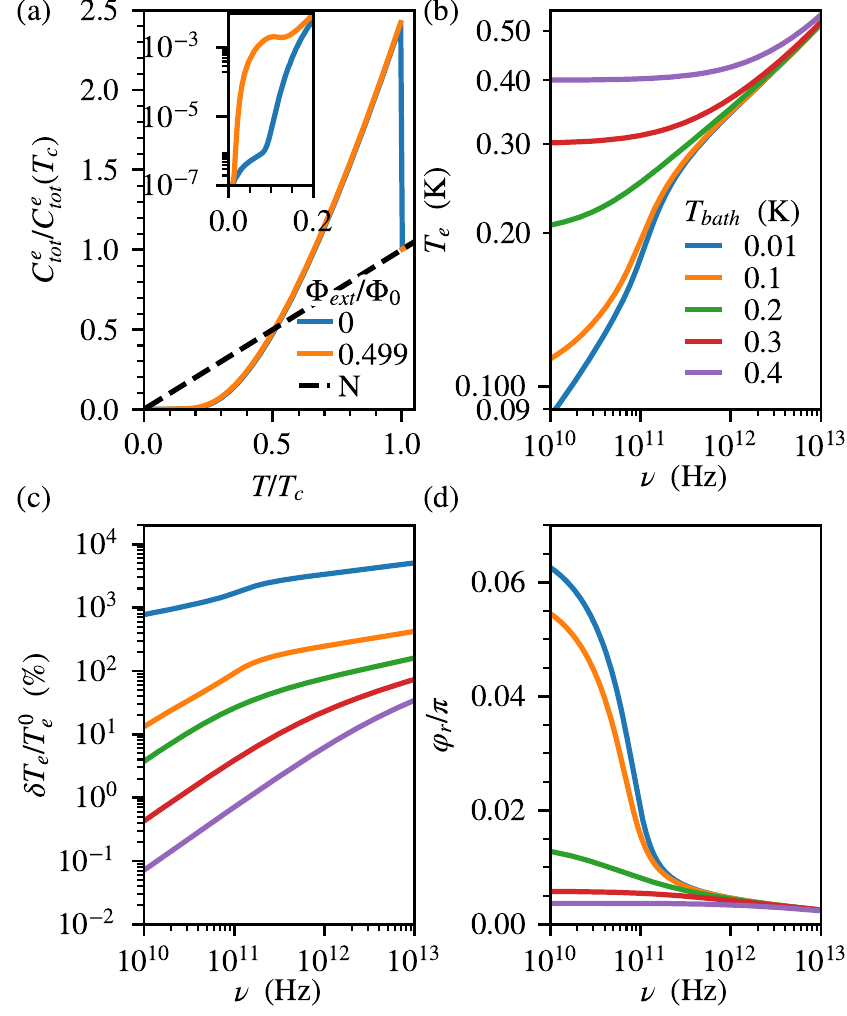}\vspace{-1mm}
  \caption{
    (a) Behavior of the total electronic heat capacity
    $C_{tot}^e$ in the detector junction vs temperature $T_e$
    calculated at 10 mK for two values of the applied magnetic
    flux. Dashed line represents the total electronic heat
    capacity when the system is in the normal state. The inset
    shows a blow-up of $C_{tot}^e$ in the low temperature regime.  
    (b) Electronic temperature in the
    detector junction $T_e$ vs $\nu$ calculated for several values of
    $T_{bath}$ at $\Phi_{ext}=0.499\Phi_0$.  (c) Relative variation of
    the electronic temperature $\delta T_e/T_e^0$ vs $\nu$ calculated
    for the same parameters as in panel (b).  (d) Phase $\varphi_r$
    across the readout weak-link vs absorbed energy $h\nu$ calculated for a
    few $T_{bath}$ values at $\Phi_{ext}=0.499\Phi_0$ and $\alpha=3$.
  }
  \label{fig:totheat}
\end{figure}

It is interesting to show first of all the behavior of the total
electronic heat capacity $C^e_{tot}$ of the detector junction, as
it determines temperature response of the sensor.  The total heat
capacity, displayed in Fig.~\ref{fig:totheat}(a), is calculated vs
temperature $T_e$ for two relevant values of the applied magnetic flux
at $T_{bath}=10$ mK.  The general behavior of $C^e_{tot}$ is the one
typical of a BCS superconductor, i.e., it is characterized by an
amplitude exponentially-suppressed with respect to that in the normal
state at low $T_e$ (i.e., for $T_e\lesssim 0.25 T_c$), and a sizable
discontinuity at the critical temperature $T_c$.  The exponential
suppression of $C^e_{tot}$ at low $T_e$ is at the origin of high
sensitivity of detection for microwave photons typical of our setup.

Figure~\ref{fig:totheat}(b) shows the final electronic temperature $T_e$ in
the detector junction as a function of energy absorbed ($h\nu$) from the
incoming photon calculated from Eq.~\eqref{heateq} for several values
of $T_{bath}$ at $\Phi=0.499\Phi_0$.  We notice, first of all, the
sizable enhancement of temperature, typically occurring below $\sim
100$ GHz, which can be achieved in the junction at low bath
temperature, in particular, for $T_{bath}\lesssim 100$ mK.  This
enhancement stems predominantly both from the exponentially-suppressed
amplitude of the junction electronic heat capacity $C^e_{tot}$ and from
the reduced volume of the N wire which is peculiar of the present
setup.  We emphasize that, for a bath temperature of 10 mK, $T_e$
obtains values as high as $\sim 90$ mK for 10 GHz photons, and up to
$\sim 150$ mK at 100 GHz.  For larger $T_{bath}$ the increase of $T_e$
is less pronounced below 100 GHz owing to temperature-driven
enhancement of the electronic heat capacity, and becomes sizable
only at frequencies exceeding 1 THz.

The relative variation of temperature, $\delta
T_e/T_e^0=[T_e(h\nu)-T_e(0)]/T_e(0)$ versus $h\nu$ for the same values
of $T_{bath}$ as in panel (b) and $\Phi=0.499\Phi_0$ is displayed in
Fig.~\ref{fig:phase}(c).  In the present setup, $\delta T_e/T_e^0$ around
$\sim 800\%$ for 10 GHz photons and around $\sim 4000\%$ for 10 THz
photons can be obtained at 10 mK of bath temperature.  By increasing
$T_{bath}$, $\delta T_e/T_e^0$ gets reduced reaching about $\sim 10\%$
for 10 GHz photons, and $\sim 300\%$ for 10 THz radiation at
$T_{bath}=100$ mK.  These results for substantial temperature
variations in the weak-link suggest that large signal to noise
ratio can be achieved with a TPC-based single-photon detector in the
microwave frequency range.

The evolution of phase $\varphi_r$ across the readout junction as a
function of photon energy $h\nu$ is shown in Fig. \ref{fig:totheat}(d) for
several bath temperatures at $\Phi=0.499\Phi_0$.  Note that at low
$T_{bath}$ the phase across the weak-link starts to be reduced already
for a few tens of GHz whereas at higher bath temperature a reduction of
$\varphi_r$ occurs only for larger frequencies.  This behavior stems
from the fact that at low bath temperature it is easy to enhance the
electron temperature also at low photon energy due to suppressed
electronic heat capacity, with the following reduction of circulating
supercurrent and thereby of phase drop across the read-out junction
[see Figs. \ref{fig:phase}(a,b)].  By contrast, at higher $T_{bath}$, the
total heat capacity is larger so that higher photon energies are
required to change appreciably the junction temperature. As a
consequence, the reduction of $\varphi_r$ is less pronounced.

%%%%%%%%%%%%%%%%%%%%%%%%%%%%%%%%%%%%%%%%%%%%%%%%%%%%%%%%%%%%%%%%%%%%%%%%%%%%%%%%%%%%%%%
\subsection{Performance: Signal to noise ratio and resolving power}
%%%%%%%%%%%%%%%%%%%%%%%%%%%%%%%%%%%%%%%%%%%%%%%%%%%%%%%%%%%%%%%%%%%%%%%%%%%%%%%%%%%%%%%

\label{sec:snperf}

In the operation as a calorimeter, i.e., in the pulsed detection mode,
we define the signal-to-noise (S/N) ratio of the detector as
\begin{equation} 
\frac{\text{S}}{\text{N}}(T_e(h\nu))=\frac{|I_p(V,T_{bath})-I_p(V,T_e(h\nu))|}{\sqrt{\mathcal{S}_{I,tot}}\sqrt{\omega}},
\label{SNratio}
\end{equation}
where $I_p(V,T_{bath})$ is current flowing through the SQUIPT biased
at voltage $V$ in the idle state, $I_p(V,T_e(h\nu))$ is the current
through the SQUIPT after the absorption of energy $h\nu$,
$\mathcal{S}_{I,tot}$ is the current noise spectral density in
the absence of radiation, and $\omega$ is the SQUIPT measurement
bandwidth.  Note that in the denominator of Eq.~\eqref{SNratio} the
current noise spectral density of the tunnel junction is evaluated in
the idle state ($h\nu=0$), where the tunneling noise contribution to
$\mathcal{S}_{I,tun}$ is larger than for $h\nu\ne0$ [see Fig. \ref{IV}(a)
and Eq. \eqref{shot}].  In the calorimeter operation mode, the
expression is appropriate for measurement bandwidths
$\omega\gtrsim2\pi/\tau$, where $\tau$ is the characteristic time
scale for the relaxation of the system back to equilibrium after
absorption of a photon.  In general, $\tau$ is determined by the
relaxation processes occurring in the weak link and, in the present
setup, the predominant energy relaxation mechanism stems from
electron-phonon interaction (see below). Moreover, we also note that
for the tunneling noise contribution, $\text{S/N}\propto 1/\sqrt{R_p}$ so
that a SQUIPT tunnel probe with lower resistance is, in general,
beneficial in order to maximize the S/N ratio.

After energy absorption, the equation governing time evolution of the
temperature $T_e$ in the sensor weak-link can be written as
\begin{equation}
C^e_{tot}(T_e,T_{bath},\Phi_{ext})\frac{d T_e}{d t}=-\dot{Q}_{e-ph}^{tot}(T_e,T_{bath}),
\label{decay}
\end{equation}
where
$\dot{Q}_{e-ph}^{tot}(T_e,T_{bath})=\dot{Q}^{N}_{e-ph}(T_e,T_{bath})+\dot{Q}^{S}_{e-ph}(T_e,T_{bath})$
is the total heat flow between electrons and lattice phonons in the
detector region.  Here, $\dot{Q}^{N}_{e-ph}$ is heat exchanged in the
N region whereas $\dot{Q}^{S}_{e-ph}$ is the one exchanged in the
lateral S portions of the detector. $\dot{Q}^{i}_{e-ph}(T_e,T_{bath})$
(with $i=N,S$) is given by
\cite{kopnin2001-ton,timofeev2009-rle}
\begin{eqnarray}
  & & \dot{Q}^i_{e-ph}=-\frac{\Sigma_i \mathcal{V}_i}{96\zeta (5)k_B^5}\int^{\infty}_{-\infty}dEE\int^{\infty}_{-\infty}d\varepsilon
  M_{E,E+\varepsilon}(\Delta_i(T_e),\Gamma_i)
  \nonumber\\
  & & \times
  \varepsilon|\varepsilon|
  [\text{coth}(\frac{\varepsilon}{2k_B T_{bath}})(f_E-f_{E+\varepsilon})-f_Ef_{E+\varepsilon}+1],
  \label{eph}
\end{eqnarray}
where $f_E(E,T_e)=\text{tanh}(\frac{E}{2k_B T_e})$,
$M^i_{E,E'}=\mathcal{N}_i^d(E)\mathcal{N}_i^d(E')-\Re[\mathcal{F}_i^d(E)\mathcal{F}_i^d(E')^*]$,
and $\Sigma_i$ is the electron-phonon coupling constant in the N(S)
region.  Here, $\mathcal{F}$ are anomalous spectral densities, which
for BCS superconductors have the form,
\begin{align}
  \mathcal{F}_S(E)
  &=
  \Delta \Re\frac{\sgn(E)}{\sqrt{(E + i\Gamma)^2 - \Delta^2}}
  \,.
\end{align}
When the structure is in the normal state, Eq.~\eqref{eph} reduces to
the well-known expression $\dot{Q}^{i,N}_{e-ph}=\Sigma_i
\mathcal{V}_i(T_e^5-T_{bath}^5)$ \cite{giazotto2006-omi}.  In our case, 
we approximate the normal-wire part of $\dot{Q}^{d}_{e-ph}$ with its normal-state value,
as for $\alpha=3$ the minigap in the detector is almost suppressed.
This choice leads to an underestimation of the S/N ratio.
We estimate numerically that the inverse proximity effect in the superconducting regions
is not important given this approximation.

The electron-phonon thermal conductances $G_{e-ph}^{N,S}$ can be
obtained by differentiating Eq.~\eqref{eph} vs. $T_e$:
\begin{align}
  \label{Geph}
  G^{i}_{th}(T)
  &=
  \frac{\partial{}\dot{Q}_{e-ph}^{i}}{\partial{}T_e}\Bigr\rvert_{T_e=T_{bath}=T}
  =
  5\Sigma_{i}{\cal V}_{i}T^4
  g_i(T)
  \,,
  \\
  \notag
  g_i(T)
  &=
  \frac{1}{960\zeta(5)}\int_{-\infty}^\infty{}
  \frac{
    dE\,d\varepsilon\;
    E|\varepsilon|^3M^i_{E,E-\varepsilon}
  }{
    k_B^6
    T^6
    \sinh\frac{\varepsilon}{2k_BT}\cosh\frac{E}{2k_BT}\cosh\frac{E-\varepsilon}{2k_BT}
  }
  \,.
\end{align}
Here, $g_i(T)$ is a dimensionless function, obtaining the value
$g_i(\infty)=1$ in the normal state or at high temperature, and
decreasing exponentially to $g(0)=(\Gamma/\Delta)^2\approx0$ at low
temperatures $k_BT\lesssim{}\Delta$ in the superconducting state.

Integration of Eq.~\eqref{decay} yields the electron-phonon relaxation
half-time,
\begin{equation}
  \tau_{1/2}(\nu,T_{bath})=\int_{[T_e^{max}(\nu)+T_{bath}]/2}^{T_e^{max}(\nu)}dT_e \frac{C^e_{tot}(T_e)}{\dot{Q}_{e-ph}^{tot}(T_e,T_{bath})},
  \label{relaxation}
\end{equation}
which allows us to determine the relevant thermal time constant of
the detector for any given energy of the incoming photon.  In
Eq.~\eqref{relaxation}, $T_e^{max}(\nu)$ is the maximum electron
temperature reached in the weak-link after absorption of a photon of
energy $h\nu$.  For small temperature changes, the heat current can be
linearized and written in terms of a thermal conductance,
$\dot{Q}^{tot}(T_e,T_{bath})\simeq(T_e-T_{bath})G_{th}^{tot}(T_{bath})$,
and in this case
$\tau_{1/2}\simeq\ln(2)C_e^{tot}(T_{bath})/G_{th}^{tot}(T_{bath})=\ln(2)\tau(T_{bath})$
is directly related to the linear thermal relaxation time
$\tau(T_{bath})$.

Finally, in the pulsed mode operation an important figure of merit is
represented by the resolving power, $h\nu/\delta E$ ($\delta E$ is the
energy resolution of full width at half maximum),
\cite{giazotto2006-omi}
\begin{equation}
\frac{h\nu}{\delta E}(\nu)=\frac{h\nu}{2\sqrt{2\ln 2}\text{NEP}_{TFN}\sqrt{\tau}},
\label{resolingpower}
\end{equation}
where $\text{NEP}_{TFN}$ is the thermal fluctuation noise limited
noise equivalent power of the sensor which stems from thermal
fluctuations between the electron and the lattice phonon system in the
detector region.  The resolving power is calculated for
$T_e=T_{bath}$, i.e., in the idle state of the detector in the absence
of radiation.  In this case, the fluctuation-dissipation theorem
states that, at the equilibrium, $\text{NEP}_{TFN}$ is given by
\cite{giazotto2006-omi}
\begin{equation}
  \text{NEP}_{TFN}=\sqrt{4k_BT^2_{bath} G^{tot}_{th}},
  \label{NEPTFN}
\end{equation}
where $G_{th}^{tot}=\partial \dot{Q}_{e-ph}^{tot}/\partial T_e$ is the total thermal conductance of the detector. As a result, recalling that $C^e_{tot}(T_{bath})=G^{tot}_{th}\tau$,
the resolving power can also  be  written as:
\begin{equation}
  \frac{h\nu}{\delta E}(\nu,T_{bath})=\frac{h\nu}{4\sqrt{2\ln 2 k_B T^2_{bath} C^e_{tot}(T_{bath})}},
  \label{resolingpower2}
\end{equation}
depending only on the electronic heat capacity.

\begin{figure}
  \includegraphics[width=\columnwidth]{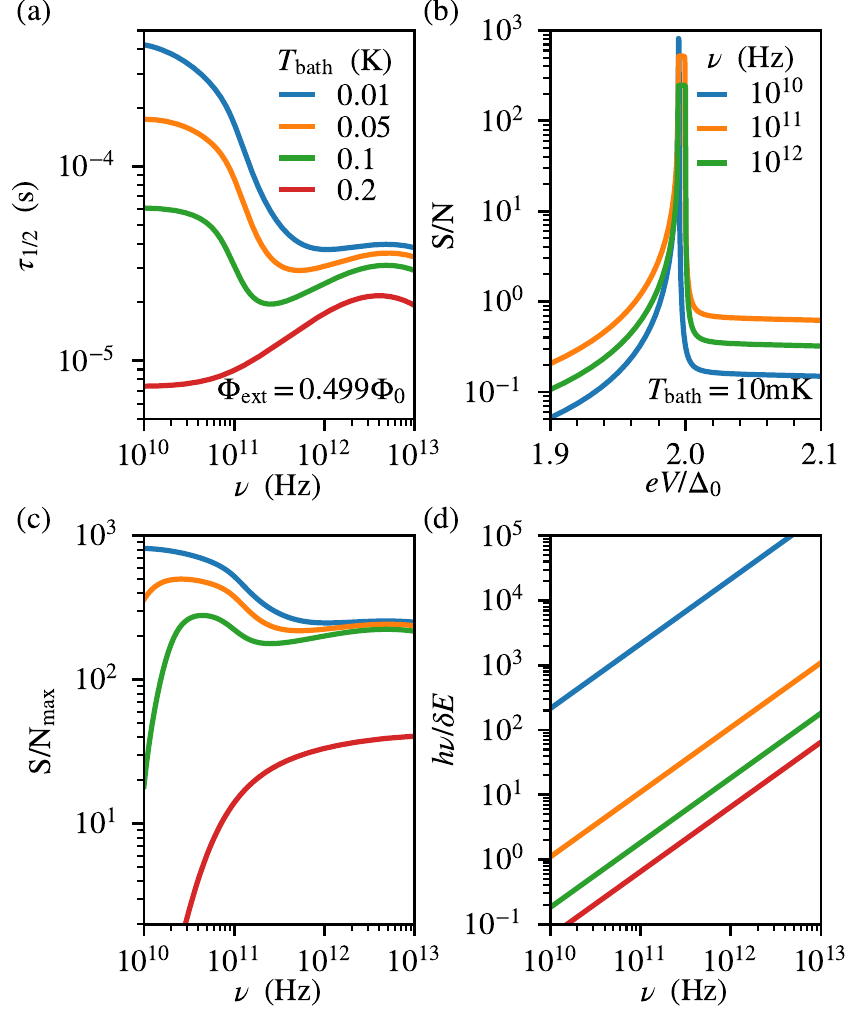}\vspace{-1mm}
  \caption{(a) Detector time constant $\tau_{1/2}$ vs absorbed energy $h\nu$ of the
    incoming radiation calculated at different bath temperatures
    $T_{bath}$.  (b) Detector signal to noise ratio (S/N) vs voltage
    $V$ calculated at 10mK for a few values of frequency of the
    incoming radiation.  (c) Maximum S/N ratio
    ($\text{S}/\text{N}_{max}$) vs $\nu$ calculated for the same bath
    temperatures as in panel (a).  (d) Detector resolving power
    $h\nu/\delta E$ vs $\nu$ calculated for the same $T_{bath}$ as in
    panel (a).  In the calculations of panels (a)-(c) we set
    $\Phi_{ext}=0.499\Phi_0$ and $\alpha=3$.
  }
\label{fig:dettime}
\end{figure}

The thermal energy fluctuations in the detector also contribute to the current
noise measured in the readout:
\begin{align}
  \mathcal{S}_{I,tot}
  &=
  \mathcal{S}_{I,tun}+\mathcal{S}_{I,TFN}
  \,,
  \\
  \mathcal{S}_{I,TFN}
  &\simeq
  |\frac{dI_p}{d(h\nu)}|^2\frac{\text{NEP}_{TFN}^2}{\tau^{-2}+\omega^2}
  \simeq
  \frac{
    |I_p(0,T_{\rm bath}) - I_p(0,T_e(\delta{}E))|^2
  }{
    \omega
  }
  \,,
\end{align}
where $\delta{}E=\text{NEP}_{TFN}/\sqrt{\omega}$ is the
root-mean-square (rms) energy fluctuation during time $t=2/\omega$ for
$\omega\gtrsim1/\tau$. The intrinsic noise $\mathcal{S}_{I,tun}$ of
the tunneling is given by Eq.~\eqref{shot}.

Figure \ref{fig:dettime}(a) shows the detector time constant $\tau_{1/2}$
calculated from Eq.~\eqref{relaxation} vs frequency $\nu$ for several
values of $T_{bath}$ at $\Phi_{ext}=0.499\Phi_0$.  From the figure it
turns out that $\tau_{1/2}\unit[\sim 10^{-5}]{s}\ldots\unit[10^{-4}]{s}$ in the range
$T_{bath}=\unit[10]{mK}\ldots\unit[100]{mK}$.

The detector signal to noise ratio S/N at $10$ mK vs bias voltage $V$
across the SQUIPT is displayed in Fig. \ref{fig:dettime}(b) for a few
values of photon energy. Here we set $\Phi_{ext}=0.499 \Phi_0$, and
assume the measurement bandwidth $\omega=2\pi/\tau_{1/2}$. We note, in
particular, that the S/N ratio is maximized close to $2\Delta_0/e$
(compare to Fig.~\ref{IV}) and, depending on $V$, it obtains values
$\gtrsim 500$.  The maximum achievable S/N ratio (S/N$_{max}$) vs
photon frequency $\nu$ is shown in Fig. \ref{fig:dettime}(c) for the
same bath temperatures as in panel (a). The S/N ratio is a
non-monotonic function of frequency, generally decreasing at
high frequency and at low frequencies where
$h\nu/\delta{}E<1$.  Notably, S/N ratios of the order of $100$ can be
obtained in the whole frequency range for bath temperatures below 100
mK. Increasing $T_{bath}$ leads to a general reduction of the S/N
ratio.

The resolving power $h\nu/\delta E$ vs photon frequency calculated for
the same $T_{bath}$ as in panel (a) is displayed in
Fig. \ref{fig:dettime}(d). In particular, the figure shows that
resolving power values $> 100$ can be achieved above 10 GHz at 10
mK. At 100 mK these values are significantly reduced due to the rapid
increase in thermal fluctuations.
The resolving power
makes the detector of potential use in \emph{microwave} and \emph{far
  infrared} single-photon detection.

\begin{figure}
  \includegraphics[width=\columnwidth]{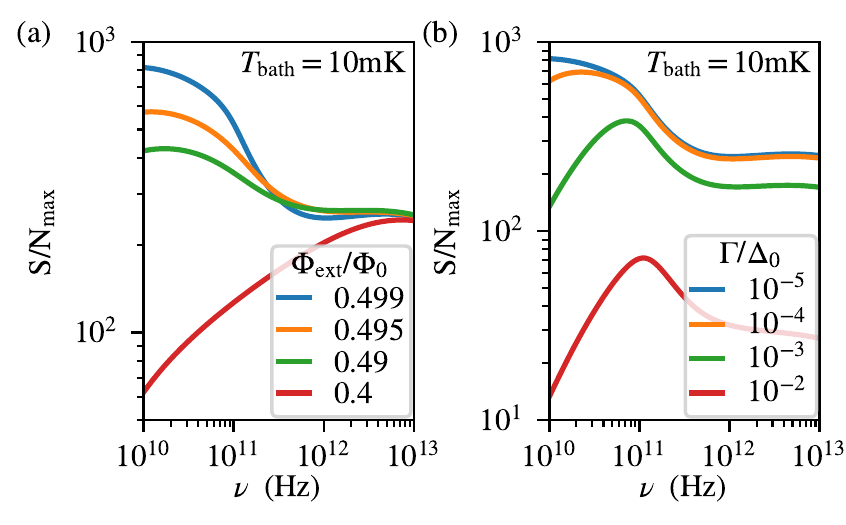}\vspace{-3mm}
  \caption{(a) Detector maximum signal to noise ratio S/N$_{max}$ vs
    frequency $\nu$ calculated for several values of $\Phi_{ext}$.
    (b) S/N$_{max}$ vs $\nu$ calculated for a few values of $\Gamma$
    at $\Phi_{ext}=0.499 \Phi_0$. 
    In all the calculations we set $T_{bath}=10$ mK.
  }
  \label{fig:detmax}
\end{figure}

The role of flux biasing of the interferometer is displayed in
Fig. \ref{fig:detmax}(a) where the S/N ratio vs frequency at 10 mK is
shown for selected values of $\Phi_{ext}$. In particular, moving away
from $\Phi=0.5\Phi_0$ leads to a reduction of the S/N ratio in the
low-frequency end, and only significant deviations lead to suppression
in the whole range.

The impact of the Dynes parameter $\Gamma$ on the signal to noise
ratio is shown in Fig. \ref{fig:detmax}(b), where S/N$_{max}$ is
calculated vs $\nu$ at 10 mK for a few values of $\Gamma$. The result
is computed using the same value in the detector and readout
junctions and in the readout probe. In particular, sufficiently small
values of $\Gamma$ have no effect ot S/N$_{max}$.  The resolving power
is similarly insensitive to it (not shown).  The effect of $\Gamma$
mainly comes from the readout junction, because under the operating
conditions in the detector junction, $\varphi^d\approx\pi$ so that the
energy gap in the detector is $\varepsilon_g^d\ll\Delta$ and there are
low-energy states contributing to heat capacity also at $\Gamma=0$.

In the above analysis, we have assumed the readout junction voltage
bias does not fluctuate. Deviation from this limit results to
photoassisted tunneling due to the fluctuations, and the convolution
\cite{ingold92} with the noise can suppress the S/N peak.  To give an
estimate of the effect, the fluctuations should be compared to the
width of the maximum of S/N, which in Fig.~\ref{fig:dettime}(b) is
$|eV-2\Delta_0|\lesssim{}10^{-2}\Delta_0$.  Modeling the noise from
the measurement circuit with an RC circuit with resistance
$R_{env}\ll{}R_p$ and parallel capacitance $C$, we have
$\delta{V}_{rms}\simeq{}\sqrt{\pi{}\omega_{\rm
    env}R_{env}^2\mathcal{S}_I}$ where $\omega_{env}=1/(R_{env}C)$
sets the maximum detector bandwidth.  For thermal noise
$\mathcal{S}_I$ from room-temperature resistance $R_{env}$,
$e\delta{V}_{rms}/\Delta_0\sim0.006(\frac{R_{env}}{\unit[100]{\Omega}})^{1/2}
(\frac{\omega_{env}}{\unit[2\pi\times10^{5}]{Hz}})^{1/2}$. The
junction shot noise can be neglected compared to the environment noise
in this case.
The effect of fluctuations can also be compared to the dependence on $\Gamma$
shown in Fig.~\ref{fig:detmax}. \cite{pekola2010-eat}
To estimate undesired signal cross-coupling from the
antenna, consider a large input power
$P_{opt}=\unit[10^{-15}]{W}=\unit[10^{13}]{Hz}\times{}h/(\unit[10^{-5}]{s})$
(c.f. the bolometer discussion in Sec.~\ref{sec:nanobolometer})
dissipated in the readout junction: the corresponding fluctuation
across the readout SNS junction is
$e\delta{}V_{rms,r}=\sqrt{R_rP_{opt}}\lesssim0.001\Delta_0$, a part of
which can contribute to the voltage fluctuations over the tunnel
probe.  We expect that with sufficient filtering, voltage stability
better than $e\delta{}V_{\mathrm{rms}}\sim10^{-2}\Delta_0$ can then be
achieved with standard room temperature electronics, which is
sufficient for the detector operation.

\section{NANOBOLOMETER}
\label{sec:nanobolometer}
%%%%%%%%%%%%%%%%%%%%%%%%%%%%%%%%%%%%%%%%%%%%%%%%%%%%%%%%%%%%%%%%%%%%%%%%%%%%%%
\subsection{Thermal model}
%%%%%%%%%%%%%%%%%%%%%%%%%%%%%%%%%%%%%%%%%%%%%%%%%%%%%%%%%%%%%%%%%%%%%%%%
\label{sec:nanobolometer-thermal}

The sensor operation in continuous power excitation (i.e., operation
as a \emph{bolometer}) can be described by considering those
mechanisms that transport energy in the N and S parts of the detector. At
low temperature, i.e., typically below 1 K, the main contribution
stems from electron-phonon heat flux which can be modeled according to
Eq.~\eqref{eph}.  In particular, the incoming radiation is first
absorbed by electrons in the weak link while the lateral contacts with
large superconducting gaps ($\Delta_1$) prevent energy from escaping from
the island. Then, the system can relax by releasing energy from
electrons to the lattice phonons residing at $T_{bath}$.  Under
absorption of a continuous power $P_{opt}$, the steady-state
temperature $T_e$ in the weak link is determined for any $T_{bath}$
from the solution of the energy balance equation for
incoming and outgoing power:
\begin{equation}
  P_{opt}+\dot{Q}_{e-ph}^{tot}(T_e,T_{bath})=0.
  \label{bolometry}
\end{equation}

\begin{figure}
  \includegraphics[width=\columnwidth]{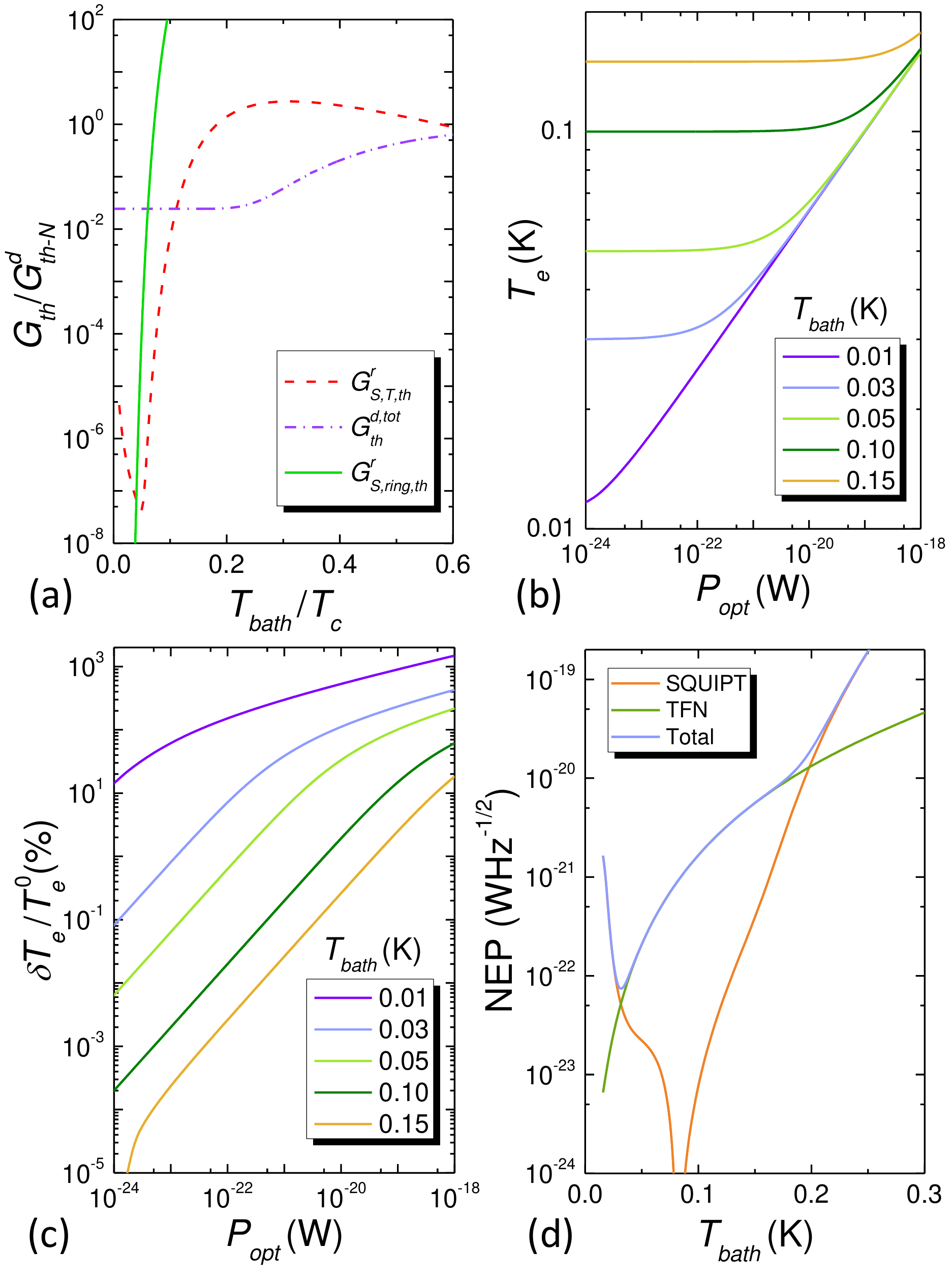}\vspace{-3mm}
  \caption{(a) Weak-link total electron-phonon thermal
    conductance $G_{th}^{d,tot}$ of the detector vs. $T_{bath}$. $G_{th-N}^d$ denotes
    the total e-ph thermal conductance in the normal state.
    The electronic heat conductivity of the SQUIPT S tunnel probe
    and ring
    ($\Gamma/\Delta_0=10^{-5}$, $G_{N,\mathrm{ring}}^r=\unit[10]{\Omega^{-1}}$)
    are also shown for comparison.
    (b) Electronic
    temperature $T_e$ in the detector junction vs optical power
    $P_{opt}$ calculated for several values of $T_{bath}$.  (c)
    Relative variation of the electronic temperature $\delta
    T_e/T_e^0$ vs $P_{opt}$ calculated for the same bath temperatures
    as in panel (c).  (d) Noise equivalent power NEP vs $T_{bath}$ for
    thermal fluctuation noise (TFN) and shot noise (SQUIPT) at $V=1.995\Delta_0$.
    In the calculations of panels (b-d) we set
    $\Phi_{ext}=0.499\Phi_0$ and $\alpha=3$. 
  }
  \label{fig:gamma}
\end{figure}

Thanks to the reduced amplitude of
$\dot{Q}_{e-ph}^{tot}(T_e,T_{bath})$ (and of the corresponding thermal
conductance $G_{th}^{tot}$) at sufficiently low temperatures,
Eq.~\eqref{bolometry} predicts that a fairly large electronic
temperature can be established in the sensor even for a quite moderate
absorbed optical power.  It is instructive to show first of all the
behavior of $G_{th}^{S}\approx{}G_{th}^{tot}$ obtained from
Eq.~\eqref{Geph} as a function of $T_{bath}$. The sensor thermal
conductance is displayed in Fig. \ref{fig:gamma}(a) (dash-dotted line). 
In particular, $G_{th}^{tot}$ turns out to
be somewhat suppressed with respect to that in the normal state
($G_{th-N}^d$), and is reduced by an order of magnitude for
$T_{bath}\lesssim0.2 T_c$. These results for
suppressed thermal conductance indicate that reduced NEP$_{TFN}$
\eqref{NEPTFN} values can be achieved with a TPC-based nanobolometer.

As anticipated above, owing to the suppression of
$G_{th}^{tot}$, a small absorbed optical power can
markedly overheat electrons in the weak link, as shown in
Fig. \ref{fig:gamma}(b). In particular, at $T_{bath}=10$ mK,
power of $10^{-22}$ W can enhance $T_e$ up to 25
mK. At a higher bath temperature the effect is less pronounced due to
enhanced electron-phonon interaction. For instance, at 50 mK $T_e$
reaches 100 mK for an input power of $\sim 10^{-19}$ W.

The relative variation of temperature, $\delta
T_e/T_e^0=[T_e(P_{opt})-T_e(P_{opt}=0)]/T_e(P_{opt}=0)$ versus
$P_{opt}$ is displayed in Fig. \ref{fig:gamma}(c) for the same values of
$T_{bath}$ as in panel (b). In the bolometric configuration, $\delta
T_e/T_e^0$ of the order of $10\%$ can be obtained for $10^{-24}$ W
and up to $300\%$ for $10^{-20}$ W at 10 mK. At $T_{bath}=50$ mK,
$\delta T_e/T_e^0$ obtains values of order $0.01\%$ and $\sim 30\%$
for the same $P_{opt}$.

We can also now comment on the assumption made above that the readout
junction remains at temperature $T\approx{}T_{bath}$.  The cooling of
the read-out weak link is provided by electronic heat transport to the
S tunnel probe, and out-diffusion to the ring.  Estimates for these
can be written as
\cite{giazotto2006-omi,bardeen1959-ttc}
\begin{align}
  \label{eq:GS-th}
  G_{S,T,th}^r
  &=
  \frac{1}{4e^2R_pk_BT^2}
  \int_{-\infty}^\infty{}dE\,
  \frac{
    E^2\mathcal{N}_N^r(E)\mathcal{N}_S^r(E)
  }{
    \cosh^2\frac{E}{2k_BT}
  }
  \,,
  \\
  G_{S,\mathrm{ring},th}^r
  &\simeq
  G_{N,\mathrm{ring},th}^r(T)
  \int_{\Delta/k_BT}^\infty
  dx\,
  \frac{
    3x^2
  }{
    2\pi^2\cosh^2\frac{x}{2}
  }
  \,,
\end{align}
where $\mathcal{N}$ are the densities of states on the two sides.  The
factor $G_{N,\mathrm{ring},th}^r(T)=L_0G_{N,\mathrm{ring}}^rT$ is the
normal-state electronic heat conductance from the junction to thermal
baths separated by the ring, where $L_0=\pi^2k_B^2/(3e^2)$ is the
Lorenz number, and $G_{N,\mathrm{ring}}^r$ a characteristic normal-state
conductance between the detector and thermal baths.  In order for the
readout to remain close to $T_{bath}$ on the relevant detector time
scales, we should have
$G_{S,th}^r=G_{S,T,th}^r+G_{S,\mathrm{ring},th}^r\gg{}G_{th}^{d,tot}$,
ie. this conductivity should be large compared to the electron-phonon
heat conductivity of the detector junction. The above conductivities
are shown in Fig.~\ref{fig:gamma}(a), together with $G_{th}^{d,tot}$,
for the parameters choices assumed above. We can observe that it
remains large compared to the electron-phonon heat conductivity
at $T_{bath}\gtrsim\unit[50]{mK}$.

Finally, we wish to comment on the thermal transport out of the detector
junction via the electron-photon heat conductance,
$G_{\mathrm{e-\gamma}}=\kappa(T)G_{Q}(T)$, where
$G_Q=\pi{}k_B^2T/(6\hbar)=G_{Q,0}T\approx{}\unit[10^{-12}]{W/K^2}\times{}T$
is the thermal conductance quantum, and $\kappa(T)$ an impedance
matching factor.  \cite{pendry1983-qlt,schwab00,timofeev2009-erq} The
matching factor $\kappa$ for BCS superconductors was considered in
Ref.~\cite{bosisio2016-phc}. We can estimate
$\kappa\sim4\Re[Z^{-1}_d]\Re[Z^{-1}_{env}]/|Z^{-1}_d+Z^{-1}_{env}|^2$
where $Z_d$ is the impedance of the detector and $Z_{env}$ that of its
total electromagnetic environment, at thermal frequencies
$\omega\lesssim{}k_BT/\hbar$.  For the detector junction at
$k_BT\sim\hbar\omega\ll\Delta$,
$Z^{-1}_d\simeq{}G_N\frac{4\Delta}{\hbar\omega}\sinh(\frac{\hbar\omega}{2k_BT})K_0(\frac{\hbar\omega}{2k_BT})e^{-\Delta/(k_BT)}-iG_N\pi\Delta/(\hbar\omega)$,
\cite{abrikosov1959-shf}
where $K_0$ is a Bessel function,
so that $\kappa(T)=\kappa_N(T)\tilde{g}(T)$,
$\tilde{g}(T)\propto{}e^{-\Delta/(k_BT)}$. In the normal state,
Ref.~\onlinecite{timofeev2009-erq} obtained $\kappa_N\sim{}10^{-3}$ for a
mismatched circuit at $\unit[100]{mK}$, which in our case corresponds
[cf. Eq.~\eqref{Geph}] to a cross-over from dominant e-ph to dominant
$\mathrm{e-\gamma}$ at the temperature
$T_{*}=[G_{Q,0}\kappa_N/(5\Sigma_d\mathcal{V}_d)]^{1/3}\times[\tilde{g}(T_*)/g(T)]^{1/3}\approx\unit[50]{mK}\times[\tilde{g}(T_*)/g(T_*)]^{1/3}$,
and $g\sim\tilde{g}$ as both scale exponentially at $k_BT\ll\Delta$ due
to the superconducting gap in the detector DOS.  Note that the
contribution of the superconducting elements and the SQUIPT readout to
$\Re[Z_{env}^{-1}]$ are similarly suppressed, so that optimizing
$G_{\mathrm{e-\gamma}}$ is likely a problem of eliminating spurious
couplings in the experimental setup.

%%%%%%%%%%%%%%%%%%%%%%%%%%%%%%%%%%%%%%%%%%%%%%%%%%%%%%%%%%%%%%%%%%%%%55
\subsection{Performance: Noise equivalent power}
%%%%%%%%%%%%%%%%%%%%%%%%%%%%%%%%%%%%%%%%%%%%%%%%%%%%%%%%%%%%%%%%%%%%%%%%%

We now turn on discussing the achievable performance of the detector
in the bolometric operation.  In this configuration, an important
figure of merit is represented by the the noise equivalent power which
stems from several uncorrelated sources of noise.  In the present
setup, the dominant contribution is due to thermal fluctuation
noise-limited noise equivalent power [NEP$_{TFN}$, see
Eq.~\eqref{NEPTFN}] whereas Johnson noise is absent owing to the
operation of the Josephson junction in the dissipationless (i.e.,
\emph{supercurrent}) regime.  The NEP$_{TFN}$ is essentially
independent of the Dynes parameter $\Gamma$.  The
contribution of the SQUIPT readout to NEP (NEP$_{SQUIPT}$)
%per measurement bandwidth
%can be obtained by solving $\frac{\text{S}}{\text{N}}(T_e(P_{opt}))=1$
%from Eq.~\eqref{SNratio} for
%$P_{opt}=\text{NEP}_{SQUIPT}\sqrt{\omega}$, taking $\mathcal{S}_{I,tot}=\mathcal{S}_{I,tun}$,
%which for $\omega\to0$ results to
is determined by $\text{NEP}_{SQUIPT}=G_{th}^{d,tot}s_T$ where $s_T$ is the temperature
sensitivity discussed in Sec.~\ref{sec:tsensitivity}.

Figure \ref{fig:gamma}(d) shows the NEP vs $T_{bath}$. The minimal total NEP
is $\approx\unit[10^{-22}]{W/\sqrt{Hz}}$ at $T_{bath}\approx{}\unit[50]{mK}$,
and is determined by the thermal fluctuation noise in the typical operation regime.

%%%%%%%%%%%%%%%%%%%%%%%%%%%%%%%%%%%%%%%%%%%%%%%%%%%%%%%%%%%%%%%%%%%%

\section{CONCLUSIONS}
\label{sec:conclusions}

Combining (i) temperature-to-phase conversion due to kinetic
inductance changes and (ii) SQUIPT tunneling spectroscopy for
detection of the variation of the phase difference provides a basis
for a thermal superconducting radiation detector.  On the temperature
sensing side, the sensitivity of the device is boosted by the strong
dependence of the SQUIPT tunneling current on the phase difference
(Fig.~\ref{IV}). Moreover, it is influenced by the effect of detector
junction kinetic inductance variation on the phase difference of the
readout junction, which is enhanced around the half-flux tuning point
$\Phi\approx\Phi_0/2$ (Fig.~\ref{fig:phase}). The thermal sensitivity
of the device is amplified by the small heat capacity of the detector
junction, both due to superconductivity (Fig.~\ref{fig:totheat}) and
the small volume, and by the reduction of electronic heat
out-diffusion by the Andreev reflections from the superconductors.  In
principle, these aims could also be achieved using longer SNS
junctions not strictly in the short-junction regime
($L\lesssim{}\xi_0$). Such a choice implies a different tradeoff
between the heat capacity (determined by volume and resistance) of the
junction, the DOS gap giving the linear-response absorption threshold,
and the kinetic inductance.  We expect lithographic fabrication of
both short \cite{ligato2017} and long \cite{giazotto2010-sqi} junction
devices is feasible.

For thermometry, the temperature-to-phase conversion differs from
e.g. NIS sensors \cite{giazotto2006-omi} in that the measurement is
nonlocal, i.e., the readout junction is separated from the detector
junction and there is little electronic thermal coupling between
them. Instead, the coupling is electrical, provided by supercurrent
flow. This is useful for improved sensitivity, as maintaining the
SQUIPT readout at a low temperature with cooling fins (here, the $S$
tunnel junction) improves its performance, and this could not be done
inside the detector junction.  We predict temperature sensitivities of
tens of $\unit{nK/\sqrt{Hz}}$ in a temperature range tunable over
$T=\unit[10]{mK}\ldots\unit[1]{K}$ with the choice of the magnetic
flux.

In the calorimetric mode, assuming that sufficiently fast measurement
of the NIS tunneling current is made, resolving power
(Fig.~\ref{fig:dettime}) of $h\nu/\delta{}E\approx1\ldots100$ is found
in the range $\nu\approx10^{10}\ldots\unit[10^{13}]{Hz}$, provided a
low bath temperature below $T_{bath}\lesssim\unit[100]{mK}$ to
suppress thermal noise.  The operation frequency range is at somewhat
lower frequencies and the resolving power higher if compared to an
inductively coupled weak link detector. \cite{giazotto2008-upj}
Compared to previously suggested small-volume superconducting
nano-HEBs \cite{karasik2011-ntp} operating in a similar temperature
range, these numbers constitute an improvement in sensitivity; compared to
Ref.~\onlinecite{wei2008-uhe}, $\delta{}E$ would be 1-2 orders of
magnitude smaller while retaining a similar thermal time constant $\tau$.

For the bolometric response, considering noise from the SQUIPT readout
and thermal fluctuations, we find NEP mostly dominated by the thermal
fluctuation noise of the detector junction for the relevant operation
temperatures. Due to the small volume and superconductivity of the
detector junction, we find
$\text{NEP}\lesssim\unit[10^{-21}]{W/\sqrt{Hz}}$ below 100 mK.  This
predicted NEP is below those obtained in recent sensitive sub-Kelvin
superconducting bolometers and kinetic inductance detectors, for which
$\text{NEP}\gtrsim\unit[10^{-20}]{W/\sqrt{Hz}}$ in similar millikelvin
temperatures have been reported. \cite{karasik2011-ntp} In comparison
to the superconducting bolometers of Ref.~\onlinecite{wei2008-uhe}, the
difference is largely due to the smaller device volume and heat
conductivity, as the limitation is due to intrinsic thermal
fluctuation noise.

In summary, we theoretically analyze a design and a readout scheme for
a superconducting weak-link radiation detector. The performance
numbers indicate an improvement over previously proposed
weak-link detectors, and would be competitive with other types of
superconducting sensors operating in similar frequency and temperature
ranges. We expect the design can be realized with current nanofabrication
technology. The detector concept can be useful for the investigation of
current open problems in astrophysics \cite{karasik2011-ntp,capparelli2016-alp} and
quantum electronic circuits.

%%%%%%%%%%%%%%%%%%%%%%%%%%%%%%%%%%%%%%%%%%%%%%%%%%%%%%%%%%%%%%%%%%%%%
\acknowledgments
%%%%%%%%%%%%%%%%%%%%%%%%%%%%%%%%%%%%%%%%%%%%%%%%%%%%%%%%%%%%%%%%%%%

F.G. and P.V. acknowledge the MIURFIRB2013-Project Coca (grant
no. RBFR1379UX) and the European Research Council under the European
Union's Seventh Framework Programme (FP7/2007-2013)/ERC grant
agreement No. 615187-COMANCHE for partial financial support.


%\bibliography{thdetector}

\end{document}